\documentclass[11pt]{ucthesis}
\usepackage{epsfig}
\def\dsp{\def\baselinestretch{2.0}\large\normalsize}
\dsp
\begin{document}

% Declarations for Front Matter

\title{Search for High Energy Neutrino Emission from Gamma-Ray Bursts with the Antarctic Muon and Neutrino Detector Array (AMANDA)}
\author{Ryan Carlton Bay}
\degreeyear{2000}
\degreesemester{Fall}
\degree{Doctor of Philosophy}
\chair{Professor P. Buford Price}
\othermembers{Professor Hitoshi Murayama\\
Professor Alexei Filippenko}
\numberofmembers{3}
\prevdegrees{B.S. (University of Wisconsin at Madison) 1994\\
M.A. (University of California at Berkeley) 1998}
\field{Physics}
\campus{Berkeley}

\maketitle
\approvalpage
\copyrightpage

\begin{abstract}
The photo-meson production of pions by shock-accelerated protons could generate a burst of $\sim10^{14}$ eV neutrinos from gamma-ray bursts (GRBs) observable in the Antarctic Muon and Neutrino Detector Array (AMANDA) or its larger successors.  Measurement of this flux can test the hypothesis that GRBs are the sources of the highest-energy cosmic rays, and GRB neutrinos could permit high-precision experiments in neutrino limiting speed, neutrino oscillations, and the weak equivalence principle.  Neutrino emission can be expected primarily during the prompt gamma-ray flash and satellite coincidence provides a well-defined window in position and time that can be searched for an excess of upgoing muon events in AMANDA from bursts in the Northern Hemisphere.  Using an event quality analysis to further reduce background in a sample of 78 GRBs from the 1997 AMANDA-B10 data set, I find a fluence limit of $E_{\nu}^2\frac{dN_\nu}{dE_\nu}<3.8\times10^{-4} {\rm min}\{1,E_\nu/E_{\rm break}\}\mbox{ TeV cm}^{-2}$ per average burst, which is orders of magnitude more stringent than in similar previous searches.
\abstractsignature
\end{abstract}

\begin{frontmatter}

\begin{dedication}
\null\vfil
{\large
\begin{center}
\emph{In loving memory of}\\
\emph{my father}\\
\emph{and}\\
\emph{Y. Franci Arzt.}\\
\emph{I miss you terribly.}\\
\end{center}}
\vfil\null
\end{dedication}

\tableofcontents
%\listoffigures
%\listoftables
\begin{acknowledgements}
To my thesis committee, Lars Bildsten, Hitoshi Murayama and Alexei Filippenko, for agreeing to oversee this foolhardy enterprise.\\
To my thesis chair, Buford Price, an ideal advisor and most amazing individual.\\
To Carrie Fordham, for hiring a listless undergraduate and giving him his first taste of real research, and whose courageous struggle against Multiple Sclerosis came to an end in 1996.\\ 
To Robert Morse, for listening to Carrie and giving me a job on this project, and who was unsuccessful in trying to talk me out of staying in the cosmic-ray business a bit longer.\\
To Francis Halzen, Eli Waxman, and Kevin Hurley for help and discussion on this fascinating little subject.\\
To Gilles Barouch and Gary Hill for their Monte Carlo calculations.\\
To John Jacobsen for helping me get this analysis off of the ground, and the sketches of nude women.\\
To Kurt Woschnagg, Predrag Miocinovic, Doug Lowder, Mike Solarz, Andrew Westphal, Ben Weaver, Dima Chirkin and Austin Richards, my Price Group brethren for continual assistance and camaraderie.\\
To Kurt Woschnagg, for advice and for allowing me to keep from really having to actually learn PAW (and to the guys in the funny shirts on the back of the manual, for reminding me that sometimes one learns by having to do it in the absolutely most difficult and roundabout way).\\
To Christian Walck and Andrew Westphal, for brain-racking discussion of the statistics of small numbers.\\
To the AMANDA collaboration, for doing so much of my work for me.\\
To the AMANDA detector.  Noisy data like this keeps you coming back, and will surely break your heart.\\
To my family, friends and those somewhere in between for love, support and affection, without which I surely would have moved to Spain to grow tomatoes (still considering it, anyway).\\
To any I have forgotten, forgive me.\\\vspace*{2cm}\\
This research was supported by the following agencies:\\ 
\hspace*{1.0cm}U.S. National Science Foundation Office of Polar Programs\\
\hspace*{1.0cm}U.S. National Science Foundation Physics Division\\
\hspace*{1.0cm}University of Wisconsin Alumni Research Foundation\\
\hspace*{1.0cm}U.S. Department of Energy\\
\hspace*{1.0cm}U.S. National Energy Research Scientific Computing Center\\
\hspace*{1.0cm}Swedish Natural Science Research Council\\
\hspace*{1.0cm}Swedish Polar Research Secretariat\\
\hspace*{1.0cm}Knut and Alice Wallenberg Foundation (Sweden)\\
\hspace*{1.0cm}German Ministry of Education and Research.
\end{acknowledgements}

\end{frontmatter}

\renewcommand{\thefootnote}{\fnsymbol{footnote}}

\chapter{Gamma-ray bursts}
\label{grb}
\setcounter{footnote}{1}

\begin{quote}
\raggedleft {\em ...and heaven will bless those stars\\that shine not the longest,\\but those that shine the brightest.}\\Anonymous 
\end{quote}

The spacecraft Vela 5B was placed roughly $180$ degrees apart from its twin, Vela 5A, at a geocentric distance of $118,000$ km on May 23, 1969.  The US satellites were part of a classified series equipped to aid in enforcement of the Partial Test Ban Treaty of 1963.  Any nuclear test detonations in the upper atmosphere or on the dark side of the moon would have generated signals in the scintillation X-ray, gamma-ray, and neutron detectors on-board.  Although no clandestine explosions were ever confirmed, during the ten-year period April 1969 to April 1979, the Vela group observed 73 instances of a very puzzling new phenomenon.  Flashes of localized gamma radiation from seemingly random directions would suddenly outshine all other sources of gamma rays combined for a few seconds and then disappear.   The discovery of gamma-ray bursts (GRBs) was announced in 1973~\cite{disco}.

The new phenomena generated tremendous excitement over the next two decades and theories of GRB origin flourished.  Some placed them at cosmological distances, others involved Galactic neutron stars, and still others reduced them to mere twinkles from the Oort cloud of our solar system.  The picture did not improve significantly until the launch of the Burst and Transient Source Experiment (BATSE) on-board the Compton Gamma-Ray Observatory in the spring of 1991.  BATSE confirmed the earlier suspicions that GRBs are very isotropic on the sky and also revealed a deficiency of faint bursts.  These facts eliminated the possibility that GRBs originate on neutron stars in the Galactic disk\footnote{The BATSE result strongly constrained but did not rule out an extended Galactic halo origin.}, and made the case for a cosmological distance scale compelling~\cite{piran, meegan}.  The science of GRBs was recently revolutionized and the cosmological hypothesis confirmed with the launch of the Italian-Dutch BeppoSAX satellite, which allowed counterpart observations to localize several GRBs to host galaxies and measure emission and absorption line redshifts of $z\approx1$~\cite{piran, sax, metz}.

\section{GRB phenomenology}
If GRBs emit their radiation isotropically, then their cosmological distance scale implies that they are extremely powerful and rare events, releasing in a matter of seconds $10^{51}$ to $10^{54}$ ergs in gamma rays alone.  Recent observations of afterglow light curve breaking~\cite{hal, kulk, harry} suggest beaming solid angles of order $\Omega\approx0.1$, which would reduce these energies by factors of $(\frac{\Omega}{4\pi})^{-1}\approx$ several hundred.  Assuming a constant formation rate, the observation of $\sim1$ per day at Earth implies a rate of only $10^{-7}(\frac{\Omega}{4\pi})^{-1}$ yr$^{-1}$ galaxy$^{-1}$.

One reason GRBs provoke such wild speculation is the enormous variety they exhibit.  In so many words, they are short flashes of low-energy gamma rays with a non-thermal spectrum.  Observed durations, however, span at least five orders of magnitude.  Light curves are complicated and vary drastically from burst to burst.  The spectrum is very hard, with emission out to GeV~\cite{hurl} and even TeV~\cite{milagro, grand} energies.  There is also now an indication of two distinct classes of GRB, with appreciably different distance scales.

\subsection{Spectra and light curves}
\label{spectrum}
\setcounter{footnote}{1}
The GRB photon spectrum can generally be well fit~\cite{band} with a double power law joined smoothly at a break energy of $\sim~200$ keV:
\[ N(\nu)=N_0 \left\{ \begin{array}
{c@{\quad \mbox{for} \quad}l}
(h\nu)^\alpha e^{(-\frac{h\nu}{E_0})} & h\nu<H \\ {[(\alpha-\beta)E_0]}^{(\alpha-\beta)}(h\nu)^\beta e^{(\beta-\alpha)} & h\nu>H\mbox{,}
\end{array} \right.  \]
where $H\equiv(\alpha-\beta)E_0$.  GRBs typically emit in the few hundred keV energy range, while prompt emission in the X-ray and lower energies is very weak.  The spectrum is exceptionally hard, and most bursts have a high-energy tail with $E^2N(E)$ nearly constant.  On the low-energy side the spectrum often follows $F_{\nu}\propto\nu^{\alpha}, -\frac{1}{2}\le\alpha\le\frac{1}{3}$, which is consistent with the tail of synchrotron emission from {\em relativistic} electrons.  The prompt gamma radiation is followed by x-ray, then optical, then radio emission over a time scale of days.  

The shortest recorded GRB lasted a mere 5 milliseconds.  The longest, GRB940217, showed GeV activity 90 minutes after the initial outburst.  GRBs 961027a, b, c, and d, the so-called ``gang of four'', occurred over a period of 2 days with overlapping error boxes.  If these are in fact the same object, then GRB durations span an astonishing seven orders of magnitude around an average of $\sim10$ seconds.  

The distribution of burst durations is bimodal with peaks at $0.3$ and $25$ seconds.  Bursts can be clearly divided into those with duration greater than and those less than $2$ seconds, with a now definite tendency for the shorter subset to have harder spectra.  There is a clear correlation between burst duration and hardness\footnote{GRB hardness is most often defined as the fluence [ergs cm$^{-2}$] in the $100-300$ keV channel divided by the fluence in the $50-100$ keV channel.} for the entire set of bursts, while the two quantities are not at all correlated within the two subsets~\cite{qin}.

Most GRBs show time structure $\delta T\ll T$, their total duration.  The shortest rise times recorded are roughly equal to the shortest structures within time histories, $\delta T_{\rm min}\approx0.2$ msec~\cite{kevin}.  This is one of our most important observational facts, since it constrains the size of the emitting region to the scale of a compact object:
\begin{center}
$R_{\rm emission}<c\delta T_{\rm min}\approx60$ km
\end{center}

\subsection{Progenitors}
All of the most fashionable theories of GRB origin eventually culminate in the formation of a black hole surrounded by a torus of debris, in order to take advantage of the potentially short time scales and large extractable energy reserve, $\sim10^{53}$ ergs, which such a system provides.  Afterglow observations of some bursts suggest dense gaseous environments characteristic of star-forming regions, giving rise to ``hypernova'' models~\cite{fryer} involving core collapse of a massive star, as explanations for the longer and softer bursts.  Mergers of binary neutron stars or neutron star-black hole systems popularly used to explain shorter GRBs are expected to involve older stellar populations in more rarefied environments, the violent formation of the binaries having propelled them to distances of several kiloparsecs from their birthplaces.  Compact binary mergers would be expected at redshifts $50\%-80\%$ those of hypernovae.  

Regardless of how the black hole-torus system forms or the mechanism through which its gravitational or spin energy is tapped, any respectable model must be able to produce sensational energy release.  When combined with observed luminosities, the size constraint above implies a region with an enormous photon energy density of about $10^{21}$ erg cm$^{-3}$.  Such a plasma has very large optical depth to pair production, yet observation requires that it be optically thin.  It was first pointed out in the 1980s by Goodman~\cite{goodman} and Paczy$\acute{\mbox{n}}$ski~\cite{pacz} that this problem can be solved if the emitting region is moving relativistically toward the observer with a sufficiently large Lorentz factor $\Gamma$.  Then the photons at the source are of energy $E_{\rm observed}/\Gamma$, and the size constraint is relieved by a factor $\Gamma^2$:
\begin{center}
$R_{\rm emission}<\Gamma^2c\delta T_{\rm min}$.
\end{center}
Infeasibly high energies are required for the {\em entire} source region to move relativistically, solely to overcome this ``compactness problem.''  The most economical scenario, both conceptually and energetically, is one in which the relativistic motion is intimately related to the observed burst.  This is achieved naturally by the relativistic fireball shock model.

\section{Great balls of fire}
\label{balls}
In the model first developed~\cite{piran} by M$\acute{\mbox{e}}$sz$\acute{\mbox{a}}$ros and Rees~\cite{rees}, Narayan et al.~\cite{naray}, and Paczy$\acute{\mbox{n}}$ski and Xu~\cite{xu}, GRBs are the result of deceleration of ultra-relativistic matter.  The model is essentially three-stage:\\
\hspace*{1cm}i) An inner engine produces a highly variable relativistic outflow of (optically thick) plasma.\\
\hspace*{1cm}ii) This energy is transported via bulk motion of the plasma out to $\sim10^{13}$ cm before it becomes optically thin.\\
\hspace*{1cm}iii) The energy is dissipated, emitting the prompt gamma rays and later radiation at longer wavelengths.\\
The plasma acceleration must saturate at a Lorentz factor $\Gamma_{\rm max}\approx \frac{E}{M_{\rm outflow}c^2}$ of at least $100$ for the gamma rays to escape pair-production losses, which constrains the mass of the flow: 
\begin{equation}
\label{mass_limit}  
{\rm M}_{\rm outflow}<10^{-5}{\rm M}_{\odot}\mbox{.}
\end{equation}
All of the energy of the fireball will be converted into bulk kinetic energy of any contamination of baryons entrained in turbulent magnetic fields.  Energy conversion is then accomplished through {\em shocks} and the observed photons are the synchrotron or synchrotron self-Compton radiation of relativistic electrons which have undergone first-order Fermi-acceleration.

The prompt gamma radiation is most likely due to {\em internal} shocks in which faster shells of the outflow overtake slower ones.  The progenitor cannot produce a single explosion but rather must generate a wind, while the burst duration and variability are determined by those of this inner engine.

{\em External} shocks also occur when the blast wave encounters the surrounding medium.  It is possible to explain the rapid variability seen in the gammas if the density of the surrounding medium is highly irregular, but this is prohibitively inefficient.  There is also no evidence that the width of features in the gamma-ray light curves increases with arrival time as the fireball decelerates.  External shocks are instead probably responsible for the afterglow at longer wavelengths over time scales of days and weeks.

\section{Future missions}
Our understanding of the nature of GRBs is certain to improve over the next several years as a new generation of satellites permits not only better measurements of the initial outburst, but also better measurements of the afterglow by an increasing number of ground-based observers.  Currently GRB afterglow observations typically begin $>>10$ seconds after the gamma-ray flash.  Fast multi-wavelength follow-up observations are needed to constrain parameters of the fireball and of the surrounding medium.  The two most notable future missions, HETE-II (High Energy Transient Explorer) and Swift, promise to provide GRB analysis in almost real time.  HETE-II will be launched in July 2000 and will quickly transmit arcminute error boxes for about $50$ bursts per year.  Swift is expected to launch in 2003 and will be equipped with a wide-field X/gamma-ray camera, a focusing soft X-ray telescope and an optical telescope.  It will quickly slew to the positions of about $300$ bursts per year for detailed X-ray and optical work~\cite{kevin}.

\chapter{Neutrino astronomy}
\begin{quote}
\raggedleft {\em Is to the sky\\
Looking to the sky\\
and down\\
Searching for a ground\\
With my good eye closed.} \\[1.5mm]
Soundgarden, {\em Searching with my good eye closed}
\end{quote}

\section{The neutrino as an experimental device}
Information about distant objects is delivered to us by light from radio frequencies up to the highest measured gamma rays {\em provided} that the photons have a clear path.  We cannot view directly radiation originating within astrophysical objects or from regions obscured by dust.  At high energies, gamma rays suffer pair-production losses on other photons whenever
\begin{center}
$4E_{\rm gamma}\epsilon_{\rm photon}>(2m_{\rm electron}c^2)^2$.
\end{center}
This means that PeV gamma-rays do not make it more than a few kiloparsecs before being absorbed by the cosmic microwave background radiation (CMBR), and TeV gammas could be absorbed within 100 megaparsecs by diffuse infrared.

Cosmic rays provide vital information as well, but are not very useful for a telescope.  A proton with energy less than about $10^{19}$ eV does not point back to its source because its direction has been randomized by intervening magnetic fields (the $\mu$G field of our Galaxy is enough).  Above this energy it will interact with the CMBR through the $\Delta$-resonance, since
\begin{center}
$4E_{\rm proton}\epsilon_{\rm photon}>(m_{\Delta}^2-m_{\rm proton}^2)c^4$.
\end{center}

Neutrinos could serve as a useful diagnostic of interesting places in the universe.  The high-energy neutrinos from the decay of mesons will trace the spectrum of those cosmic rays interacting either with interstellar gas or possibly even at their acceleration sites.  High-energy neutrino production is also expected in the decay or annihilation of ultra-heavy particles.  Such particles might be concentrated at the core of the Earth, Sun, or Galactic center, where relic weakly interacting massive particles (WIMPs) produced during the Big Bang have fallen into a gravity well and been trapped via elastic scattering.

The neutrino can also escape its birthplace and traverse vast distances unimpeded.  With its tiny mass and inability to take part in any but the weak interactions, the interaction length of a $1$ TeV neutrino, for example, is about 2.5 million kilometers of water~\cite{gandhi}.  Unfortunately these same properties that make the neutrino an immutable courier of information make it diabolically elusive to detection.  Nonetheless, the neutrino telescope can look to several experimental precedents for encouragement.  Solar neutrinos have been observed both radiochemically~\cite{sun} and through the observation of recoiling electrons~\cite{sno}.  Low-energy neutrino astronomy was christened with the detection of neutrinos from supernova SN1987A~\cite{kam,imb}.  More recently, particle physics using atmospheric neutrinos has become a serious science, with convincing evidence for neutrino oscillations~\cite{superk}.

\section{Detection method}
With cross-sections around $10^{-(\mbox{thirty-something})}$ cm$^2$, any experiment bent on neutrino detection must patiently await and somehow discern the particle's rare interactions.  In one conspicuous reaction of great interest for the purposes of a neutrino telescope, the $\nu_\mu$ produces a muon through the charged-current interaction.  This daughter muon carries away one-third to one-half of the neutrino energy, and at high energies preserves the neutrino direction with a mean deviation of
\[\sqrt{\langle \theta_{\mu\nu}^2 \rangle}[{\rm radians}] \approx \sqrt{\frac{m_{\rm p}}{E_\nu}}\mbox{,}\]
which is about one degree at $3$ TeV~\cite{gaisser}.  This muon can be traced by the $\sim250$ $\check{\rm C}$erenkov photons it emits per centimeter as it traverses a transparent medium embedded with a lattice of photomultiplier tubes (PMTs)~\cite{gaisser}.  Unlike other much smaller water $\check{\rm C}$erenkov detectors which have typically a factor of $10^3$ higher density of PMTs, the neutrino telescope is designed primarily for the TeV range, taking advantage of the long lever arm furnished by a high-energy muon track.  A muon in ice or water loses only about $2$ MeV cm$^{-1}$ to ionization up to $5$ TeV, above which the energy loss rises approximately linearly with energy due to pair production, bremsstrahlung, and nuclear interactions.  A muon which does not suffer catastrophic energy loss can reach the instrumented volume of a detector several kilometers away from a high-energy neutrino interaction within a large effective volume.  The electromagnetic showers induced along the tracks of the higher-energy muons add to their light generation, increasing detector effective area with energy as well.  Calorimetry is possible albeit imprecise, and depends strongly on PMT spacing and the optical properties of the medium.
%\begin{figure}[h]
%\begin{center}
%\epsfig{file=method.eps,height=3.4in}
%\caption{\label{method} A relativistic muon can be traced by its $\check{\rm C}$erenkov cone as it traverses a lattice of photomultiplier tubes embedded in a transparent medium.}
%\end{center}
%\end{figure}
\begin{figure}[h]
\begin{center}
\epsfig{file=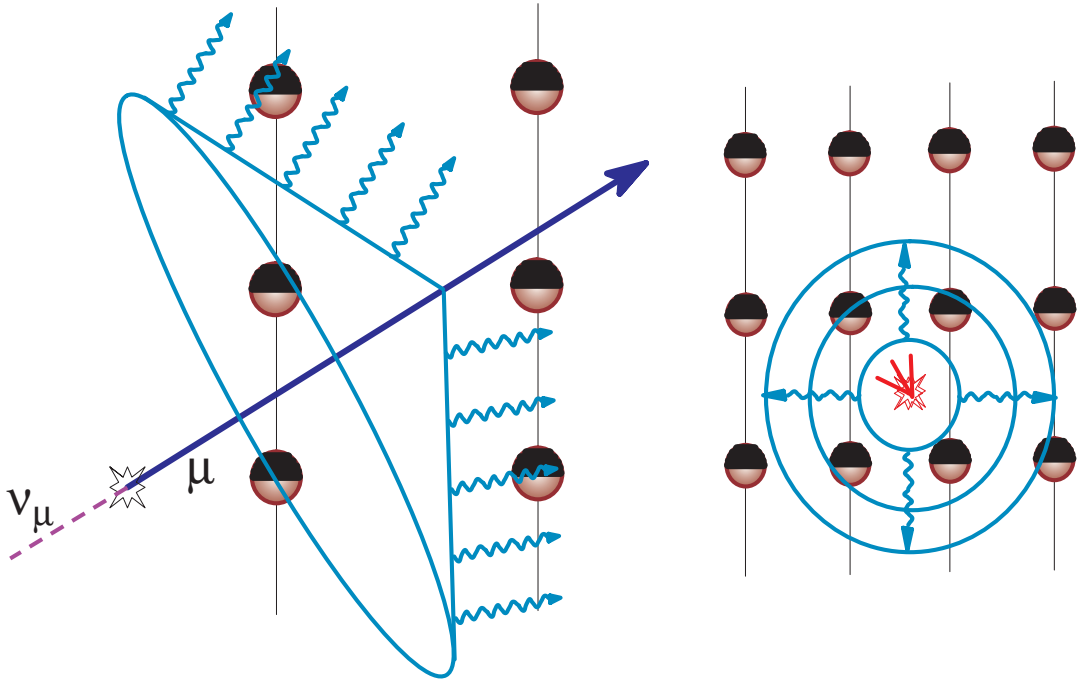,height=3.6in}
\caption{\label{method} ({\em Left}) A relativistic muon can be traced by its $\check{\rm C}$erenkov cone as it traverses a lattice of photomultiplier tubes.  ({\em Right}) An electromagnetic or hadronic shower event.}
\end{center}
\end{figure}

A lattice of PMTs provides other modes of neutrino detection.  Reconstruction of the energy and perhaps even direction of contained cascade events arising from electron neutrinos, stopping muons and the ``double bang'' signature of a tau neutrino is possible.  The low-energy neutrino burst from a nearby supernova can be noticed by carefully monitoring only the counting rate of the PMTs~\cite{ralf}.

At low energy thresholds a neutrino telescope must cope with the comparatively enormous flux of muons produced in the atmosphere overhead.  The simplest strategy is to accept only upward traveling muons:  an upgoing muon is certain to have been a neutrino which penetrated the filter of the Earth.  The detector must then be able to distinguish an upward- from downward-traveling event very reliably.  An above-ground detector would need better than $10^{11}$ discrimination at sea level, while a detector shielded under 1000 meters of water still needs about $10^6$.
\begin{figure}[h]
\begin{center}
\epsfig{file=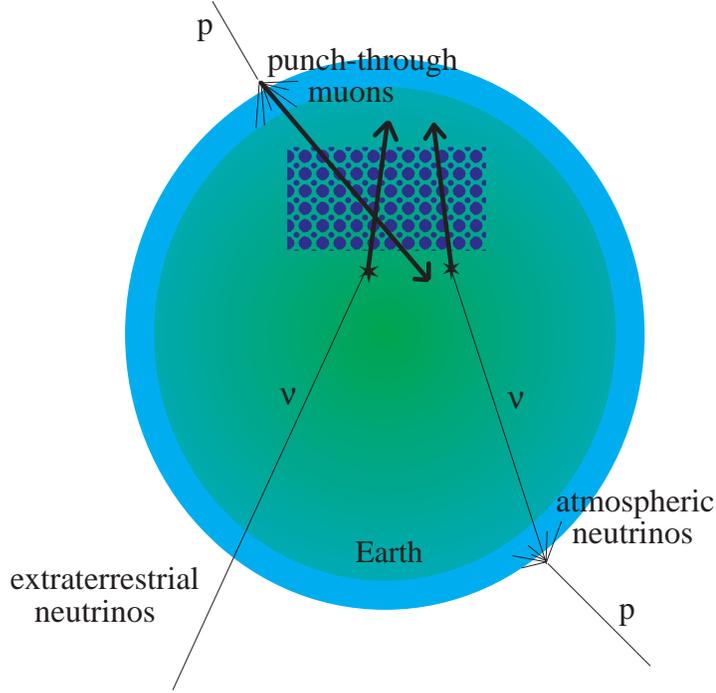,height=3.6in}
\caption{\label{earth} The neutrino telescope must determine muon direction reliably in order to distinguish extraterrestrial neutrinos from atmospheric backgrounds.}
\end{center}
\end{figure}

Neutrinos are also produced in cascades from particle collisions in the atmosphere on the other side of the Earth.  In general the flux of neutrinos from pion decay is given by~\cite{gaisser}
\[\frac{dN_\nu}{dE_\nu}=\frac{N_0(E_\nu)}{1-Z_{NN}}\left\{ \frac{A_{\pi\nu}}{1+B_{\pi\nu}\cos{\theta}E_\nu/\epsilon_\pi}+(\cdot\cdot\cdot) \right\}\mbox{,} \]
where $A_{\pi\nu}=Z_{N\pi}(1-r_\pi)^\gamma/(\gamma+1)$, $r_\pi=(m_\mu/m_\pi)^2$, $B_{\pi\nu}$ is a constant that depends on the spectral index $\gamma$ and nucleon and pion attenuation lengths, and the $(\cdot\cdot\cdot)$ term represents the contribution from other mesons.  The energy $\epsilon_\pi$ is a characteristic quantity that reflects the competition between decay and interaction in the medium, and for cascade development in the Earth's atmosphere $\epsilon_\pi\approx115$ GeV.  Although there is no direct way to distinguish these from the extraterrestrial variety, atmospheric neutrinos have a smooth angular distribution and above several TeV their spectrum falls precipitously as $E^{-3.8}$, at least out to an energy above which charm production (with a flatter spectrum) dominates.  In investigations of localized transient phenomena, background from atmospheric neutrinos is virtually negligible.
\begin{figure}[h]
\begin{center}
\epsfig{file=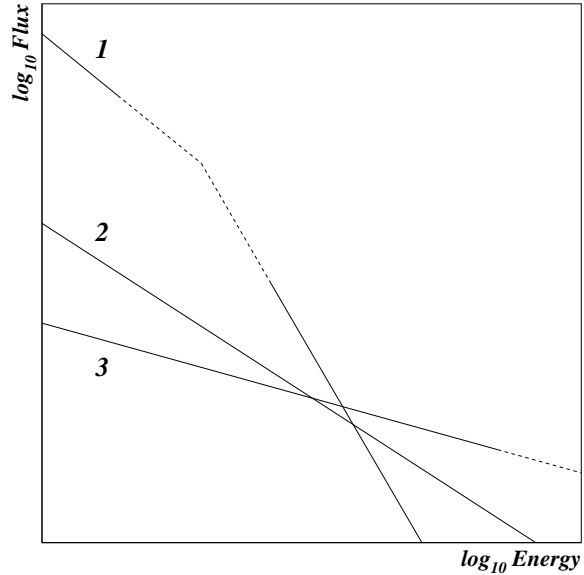,height=3.0in}
\caption{\label{kluge} Schematic presentation~\cite{gaisser} of neutrino fluxes from (1) atmospheric neutrinos, (2) cosmic-ray interactions with interstellar gas and (3) cosmic-ray interactions at their source.}
\end{center}
\end{figure}

The $\nu_\mu + A \rightarrow \mu + X$ cross-section also grows linearly with energy up to about $10^{14}$ eV and then levels off to increase as $E_{\nu}^{0.4}$ when the mass of the {\it W} propagator begins to impede its coupling.  Neutrino absorption in the Earth increases until eventually looking downward through the planet becomes a disadvantage.  Above about 40 TeV, the neutrino interaction length becomes smaller than an Earth diameter, about $10^8$ m of water equivalent.  The stiffer the neutrino spectrum, the more events are confined to solid angle closest to the horizon.

\chapter{AMANDA}
The Antarctic Muon and Neutrino Detector Array is an American and European effort to construct a neutrino telescope at South Pole Station using the Antarctic ice cap~\cite{og_2}.  This ice is a stable platform for the delicate instruments of a particle experiment, and also an extremely transparent optical medium with manageable scattering for faint light detection with sparsely distributed PMTs.  The unique location of AMANDA permits (and in some cases \textit{requires}) study in other scientific disciplines including glaciology, seismology, paleoclimatology, and even microbiology.  

Construction of the 10-string phase of the array, AMANDA-B10, was completed in the austral summer 1996-97 and its geometry and calibration established in 1998.  AMANDA-B10 has a threshold of about 50 GeV and achieves an angular resolution of 2.5 degrees, while presenting an effective area on the order of $10,000$ m$^2$ to the highest-energy neutrinos.  The data set from the 1997 season of this first-stage prototype will be the primary subject of this analysis and is relatively well understood.  Significant progress into calibration of the 1998 and 1999 data has been made.

\section{Strings of beads}
The B10 array consists of 302 PMTs arranged on 10 strings at depths between 1.5 and 2 km in approximately a right cylinder of 60 meter radius, a geometry biased to vertically-traveling muons.  The acceptance of the array was made more symmetric with the addition of three strings in 1997-98 and six more strings in 1999-00 for a total of 19 strings and a 90 meter radius.  This AMANDA-II will comprise the core of the IceCube kilometer-scale array still in the proposal stage at the time of this writing~\cite{icecube}.

The maximum construction rate of the array achieved thus far has been 6 strings per season.  A large hot-water drill is used to make a $60$ cm diameter wet hole down to a depth of 2 km, and the AMANDA string is then deployed before the hole refreezes.  Complete refreeze typically takes several days due to residual heat.  
\begin{figure}[p]
\begin{center}
\epsfig{file=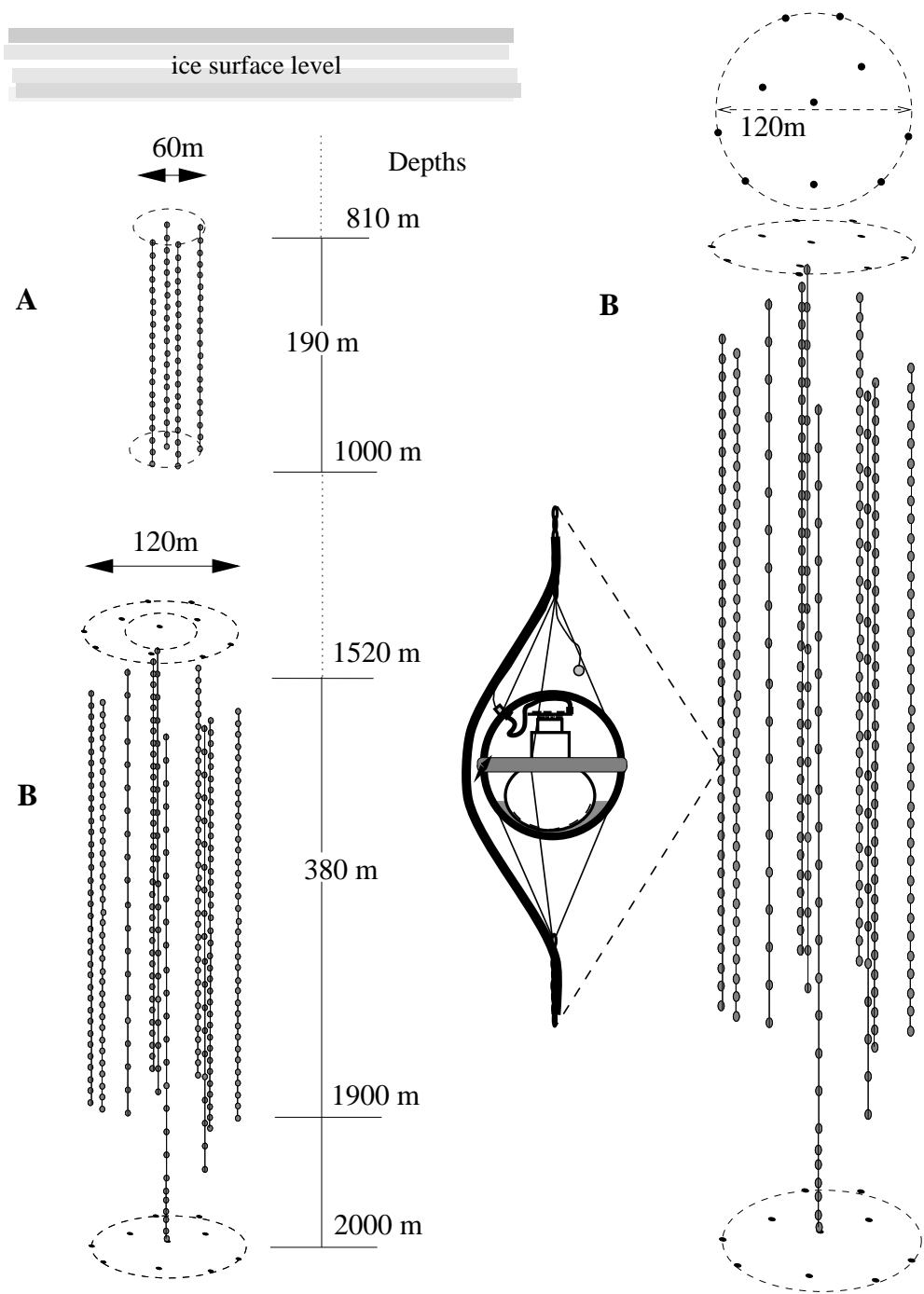, height=7.2in}
\caption{\label{array} The first four strings were deployed too shallow in ice contaminated by air bubbles which have not yet converted into clathrates (air-hydrate crystals), and these comprise AMANDA-A.  AMANDA-B was deployed much deeper in bubble-free ice.  An individual optical module is exploded to illustrate its mechanical design.}
\end{center}
\end{figure}
An optical module (OM) consists of an 8-inch PMT and base within a high-pressure glass sphere, with its own electrical connection to the surface for power and signal transmission, and a fiber optic connection for calibration purposes.  Signal processing is carried out entirely on the surface in the 10-string prototype.  Although this configuration is quite robust and the low background rates of PMTs in ice allow for a very simple triggering scheme, the collaboration is exploring many more sophisticated technologies for use in the IceCube array.  Fiber optic signal transmission and full  waveform digitization can improve photoelectron timing and dynamic range.  Local coincidence triggering between neighboring OMs might become necessary in place of the simple multiplicity trigger as the number of OMs continues to grow~\cite{nygren}.

\section{An all-natural optical medium}
The southern polar ice cap is the product of hundreds of thousands of years of steady accumulation of less than a foot per year of very finely powdered snow.  Ground-penetrating radar isochron maps taken from airplanes suggest that, on the kilometer scale, South Pole ice can primarily be described as a function of depth.  Layers of higher concentrations of various kinds of dust correspond to glacial maxima (ice ages), when the atmosphere was more arid and the larger temperature gradient between the poles and the equator created higher winds.  

AMANDA has made possible the study of the optical properties of South Pole ice at depths from 800 to 2300 meters~\cite{woschnagg, ask95, ask97, amanda97, berg, he}.  Isotropized and  filtered light from pulsed LEDs and lasers, DC LEDs, halogen lamps, and xenon arc lamps deployed alongside or inside the optical modules, and laser pulses transmitted via fiber optic from the surface serve as beacons to OMs as far as 200 meters away.  

A diffusive analytic solution can be applied effectively only when the source and receiver are separated by a distance $d$ much larger than one effective scattering length, $\lambda_{\rm e}\equiv \lambda_{\rm scat}/(1-\langle\cos\theta\rangle)$.  If the scattering length is much larger than the spacing (the ideal case for a telescope), delay time is a delta function.  Monte Carlo simulation becomes necessary to describe bulk properties when $d\approx\lambda_{\rm e}$, as is the case at depths $\ge 1400$ meters in AMANDA.  Pulsed sources can be used to extract separately both the effective scattering length and absorption length, $\lambda_{\rm a}$, from arrival-time distributions by fit to simulation.  The propagation length $\lambda_{\rm prop}\equiv\sqrt{\lambda_{\rm a}\lambda_{\rm e}/3}$ can be determined over large distances from a measurement of fluence versus distance falling as $N(d)\propto\frac{1}{d}e^{-(\frac{d}{\lambda_{\rm prop}})}$.  

\begin{figure}[p]
\begin{center}
\epsfig{file=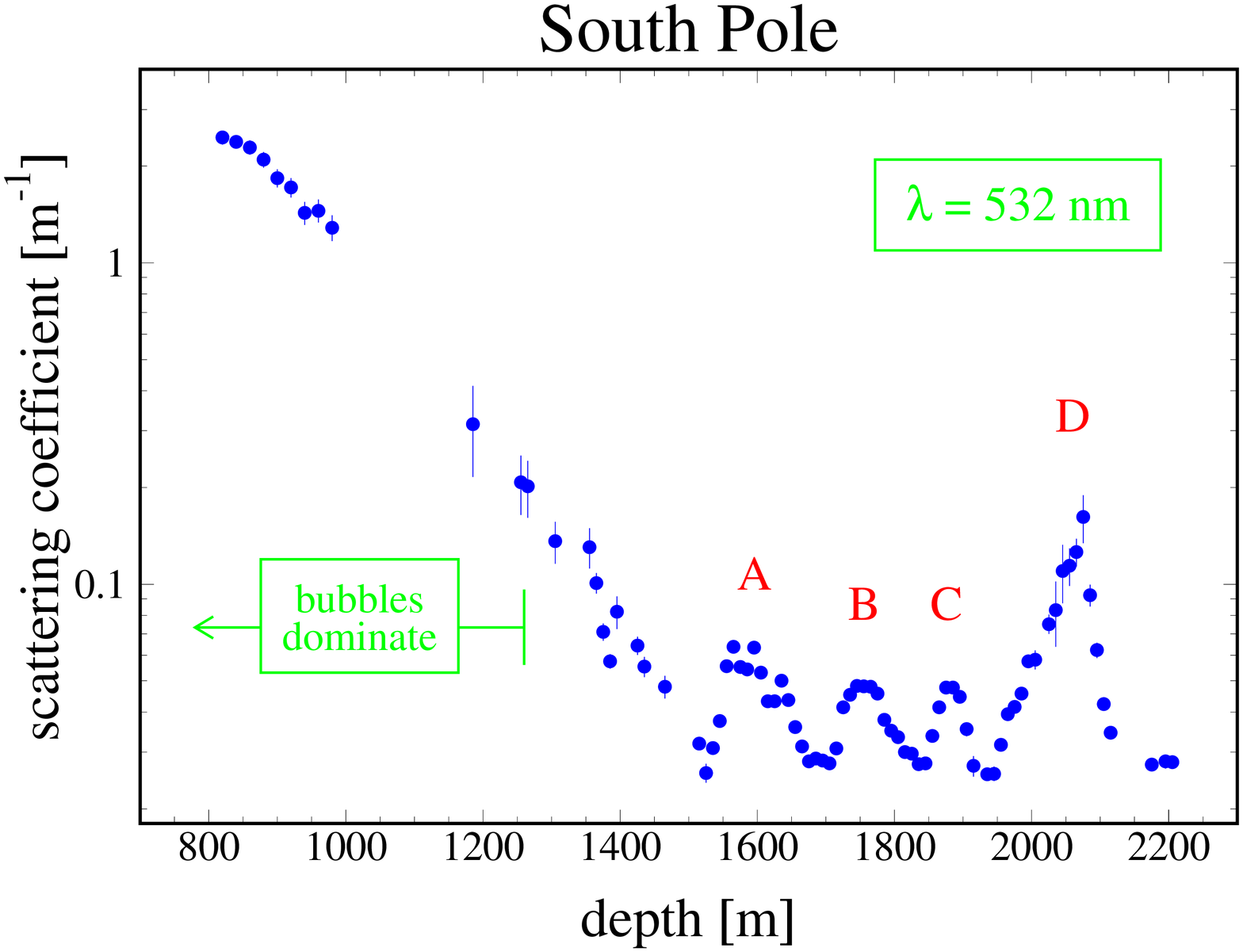, height=3.6in}
\epsfig{file=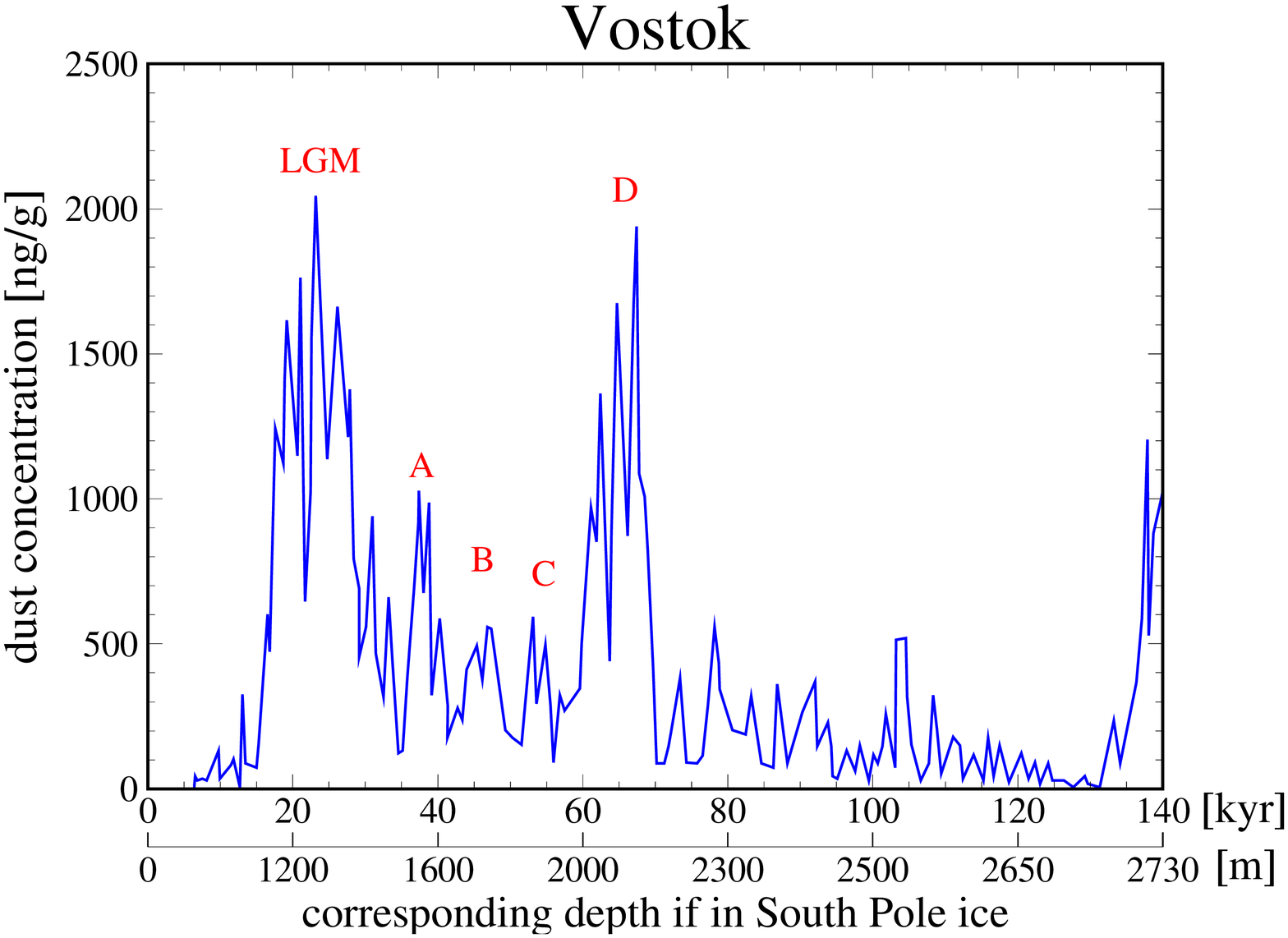, height=3.8in}
\caption{\label{age} Matching the scattering coefficient measured optically at South Pole to ice core data at Vostok station determines the relation between age and depth.  LGM is Last Glacial Maximum, while D signifies a lesser cold period not generally considered an ice age.  A, B, and C are still smaller anomalies used as reference points.}
\end{center}
\end{figure}

Emitter/receiver combinations in the same horizontal plane at as many depths as possible are desirable to avoid integrating over vertically varying features such as dust bands.  Optical properties can be predicted all the way down to bedrock by matching variations measured with optical methods to dust concentration profiles of ice cores taken at the Russian Vostok site $\sim1300$ km away and the Japanese Dome Fuji site $\sim1500$ km away and extrapolating (Figure~\ref{age})~\cite{kurt}.  More is now understood about the optical properties of the Antarctic ice cap than is known about those of the deep ocean~\cite{buf}.

\begin{figure}[h]
\begin{center}
\epsfig{file=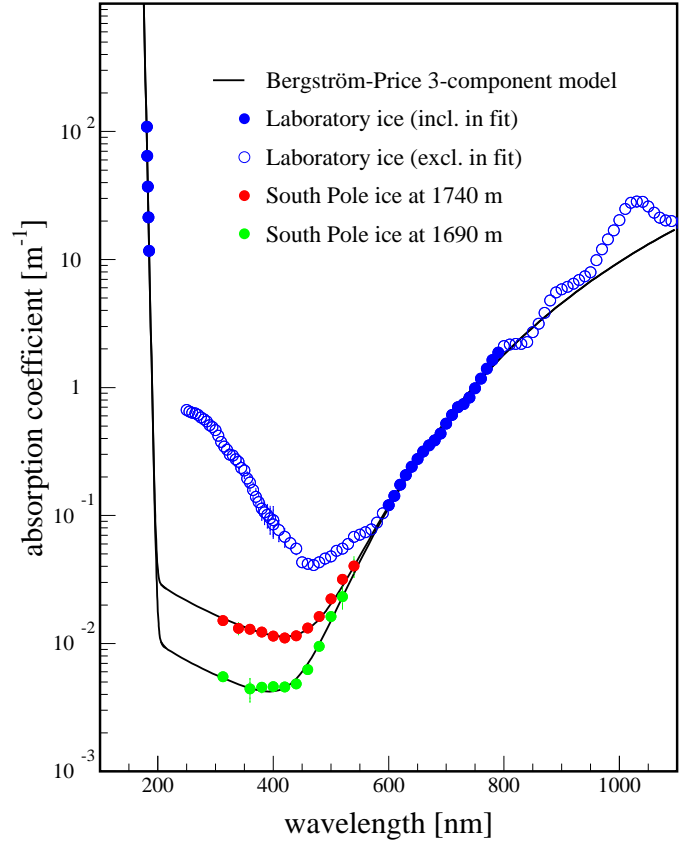, height=4.4in}
\caption{\label{3com} The three-component model of South Pole ice absorption.  In the visible window between $4\times10^{14}$ Hz and $8\times10^{14}$ Hz the transparency is limited only by a small level of dust contamination.}
\end{center}
\end{figure}

\begin{figure}[p]
\begin{center}
\epsfig{file=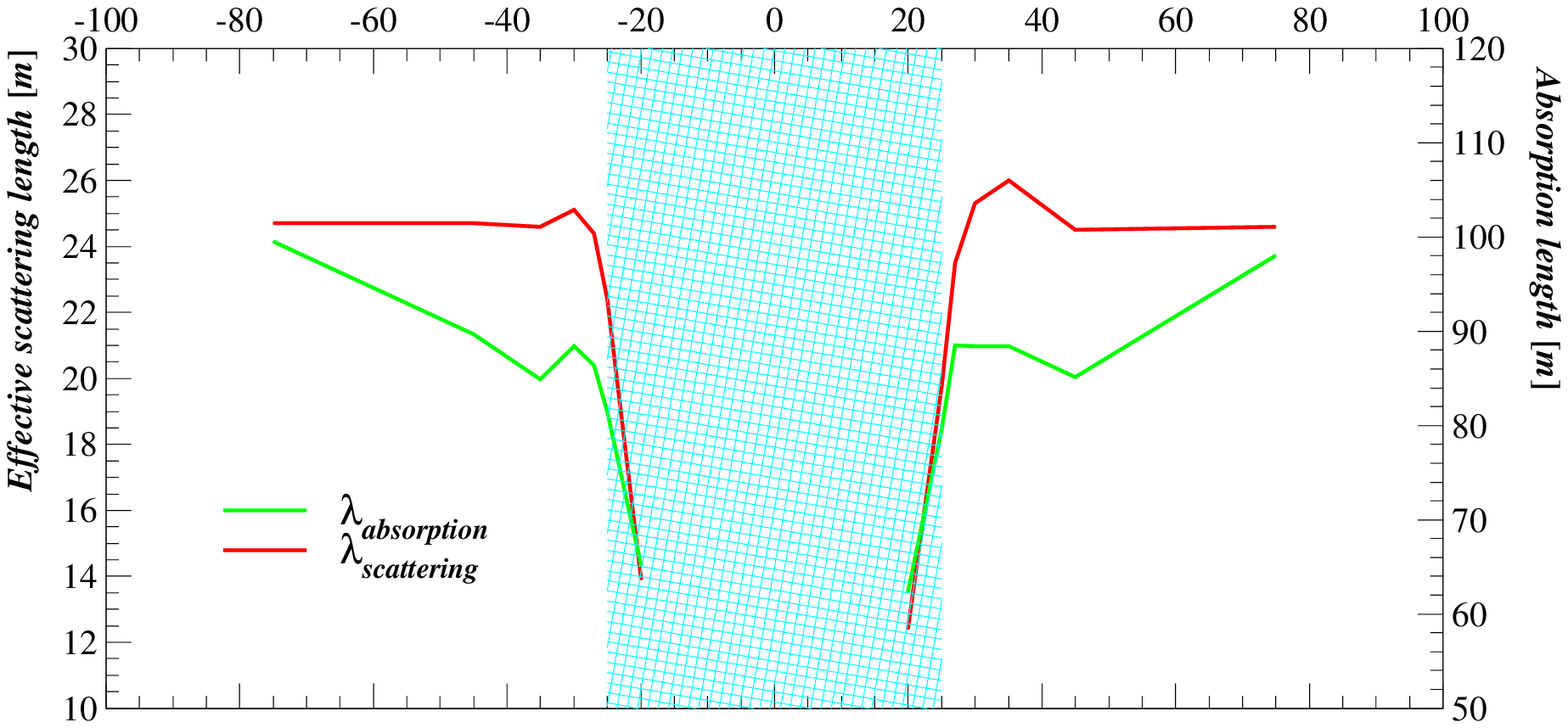, height=2in}
\epsfig{file=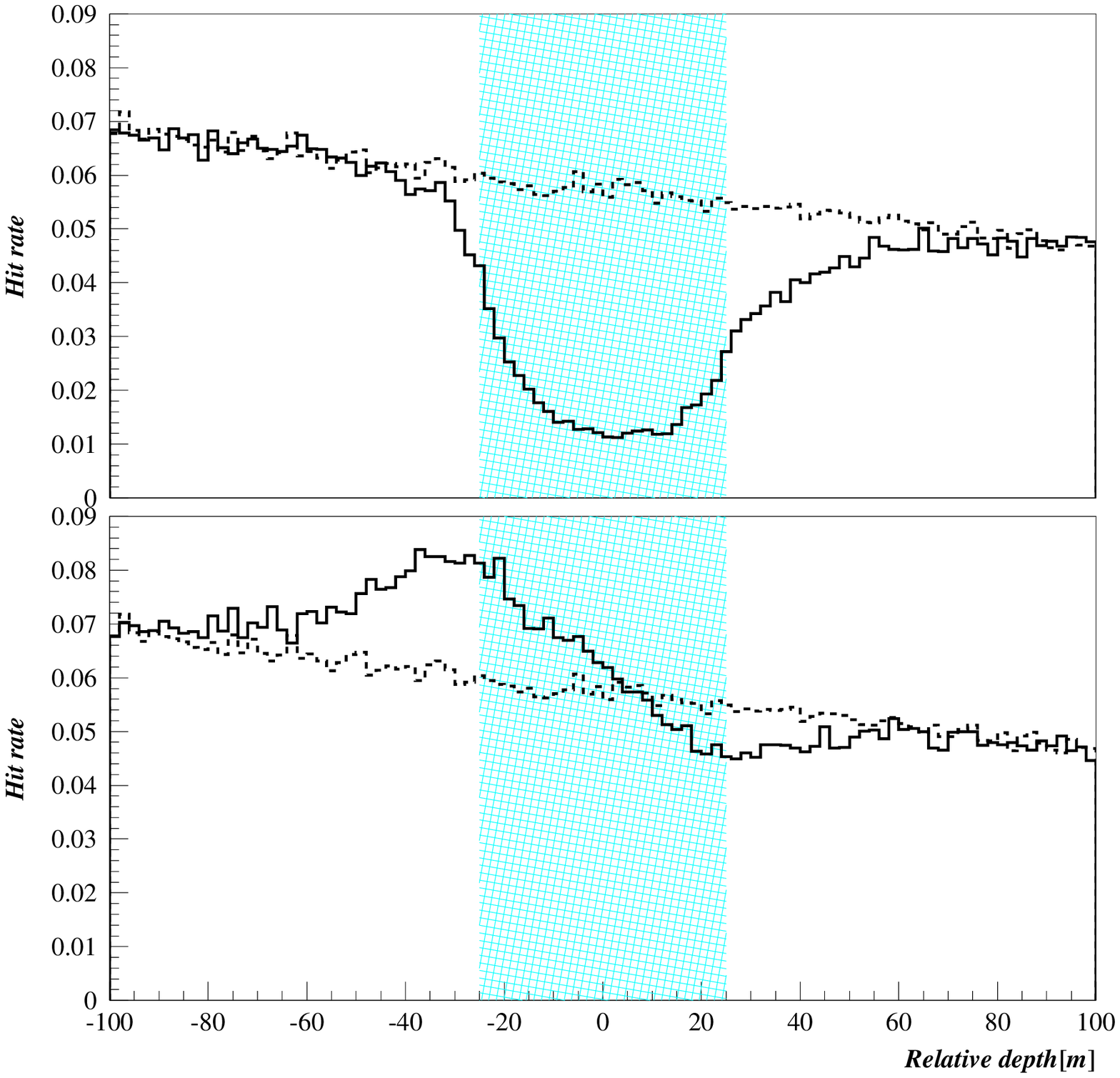, height=4.6in}
\caption{\label{layer} ({\em Top}) Monte Carlo simulation in which optical properties are fitted to an arrival time distribution between an emitter and receiver separated by 50 meters at varying distance from a dust layer with three times more scattering and absorption.  Near the layer, optical properties are actually perceived to {\em improve}, due to the layer acting as a backboard.  ({\em Middle}) Monte Carlo simulation of the response of an array of optical modules to downgoing muons when an impurity layer is introduced with three times more absorption and scattering and ({\em bottom}) with increased scattering only.  The dashed lines show response with uniform ice.  In the scattering-only case the layer acts as a photon guide and enhances the hit rate on the upstream side.}
\end{center}
\end{figure}

Figure~\ref{3com} is a plot of South Pole ice absorption.  Shown are data fitted with a ``three-component'' model of Bergstr$\ddot{\mbox{o}}$m and Price~\cite{berg}.  Absorption increases dramatically at wavelengths shorter than the visible due to electronic excitation modes and at wavelengths longer than the visible due to molecular modes.  Inside this window the transparency of South Pole ice is limited only by the small concentrations of dust, with an absorption length on the order of $100$ meters.  This is many times more transparent than laboratory ice, and combined with the moderate concentration of scattering impurities which act as wave-guides, South Pole ice provides a natural medium for efficient calorimetry of electromagnetic processes.
\newpage
The simulation~\cite{og} in Figure~\ref{layer} demonstrates how the response of a $\check{\rm C}$erenkov detector can be affected when optical properties are not uniform.  Layers of enhanced concentrations of various impurities can either decrease or increase the light-collection efficiency of the array.  The effective scattering length is found to vary by a factor of two over AMANDA-B10, yet all currently distributed collaboration software (including all software used in this analysis) assumes uniform optical properties for the sake of simplicity and speed.  This is a primary cause of current disagreement between data and Monte Carlo simulations, and introduces considerable uncertainty into any estimation of sensitivity.

%\begin{figure}[p]
%\begin{center}
%\epsfig{file=layer2.eps, height=6.0in}
%\caption{\label{layer} Monte Carlo of the response of an array of optical modules to downgoing muons when a 50 meter impurity layer is introduced ({\em top}) with three times more absorption and scattering and ({\em bottom}) with increased scattering only.  The dashed lines show response with uniform ice.  Note in the scattering-only case the layer acts as a photon guide and actually enhances hit rate on the upstream side.}
%\end{center}
%\end{figure}
%\begin{figure}[p]
%\begin{center}
%\epsfig{file=bump.eps, height=6.0in}
%\caption{\label{bump} Monte Carlo in which optical properties are fitted by placing an emitter and receiver separated by 50 meters at varying distance from a dust layer with three times more scattering and absorption.  Near the layer, optical properties are actually perceived to {\em improve}, due to the layer acting as a backboard.}
%\end{center}
%\end{figure}
\section{Event reconstruction}
\begin{figure}[h]
\begin{center}
\epsfig{file=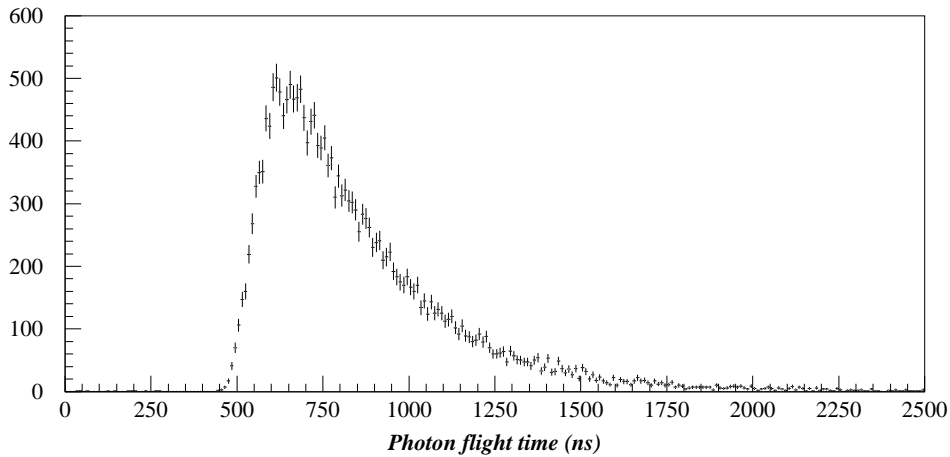, height=2.5in}
\caption{\label{timing} The arrival time distribution of a pulsed blue LED at a distance of 100 meters.}
\end{center}
\end{figure}

The arrival time distribution of a population of photons emitted from a pulsed point source is a combination of delay due to scattering impurities and the usual Gaussian timing uncertainties of several nanoseconds due to electronics (Figure \ref{timing}).
Because a neutrino telescope is sparsely instrumented and AMANDA-B10 is still small in comparison to the range of the muons it is designed to detect, a likelihood analysis is necessary instead of a simple ${\chi}^2$ minimization to make the most of degraded information delivered by delayed photons.  The quantity $p_{i}(t)$ gives the probability of OM $i$ being hit at time $t$, and reconstruction seeks to maximize~\cite{wieb_reco}
\begin{center}
$\log(\mathcal{L}) = \log(\prod_{\rm all hits} p_i) = \sum_{\rm all hits} \log(p_i)$.
\end{center}

\begin{figure}[h]
\begin{center}
\epsfig{file=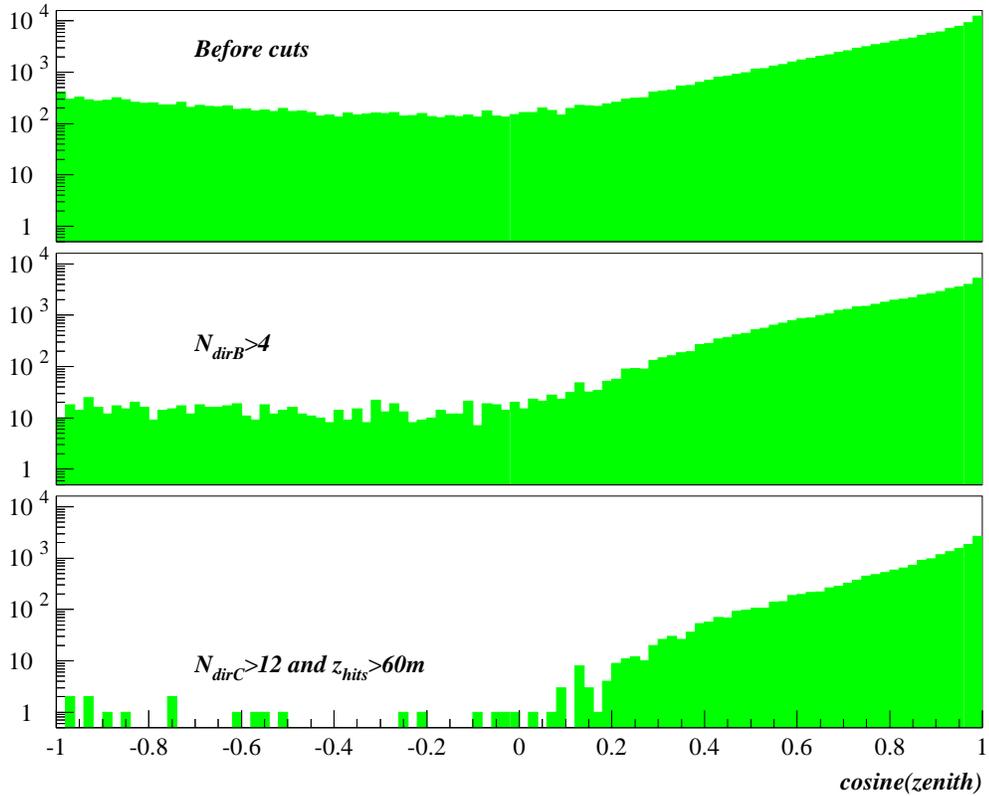, height=4.2in}
\caption{\label{zenith} Reconstructed zenith angle at three cut levels on the same sample of events.  Requiring a minimum number of direct hits or, e.g., a minimum span of hits along a track ($z_{\rm hits}$) removes events misreconstructed as upgoing.}
\end{center}
\end{figure}

\begin{figure}[p]
\begin{center}
\epsfig{file=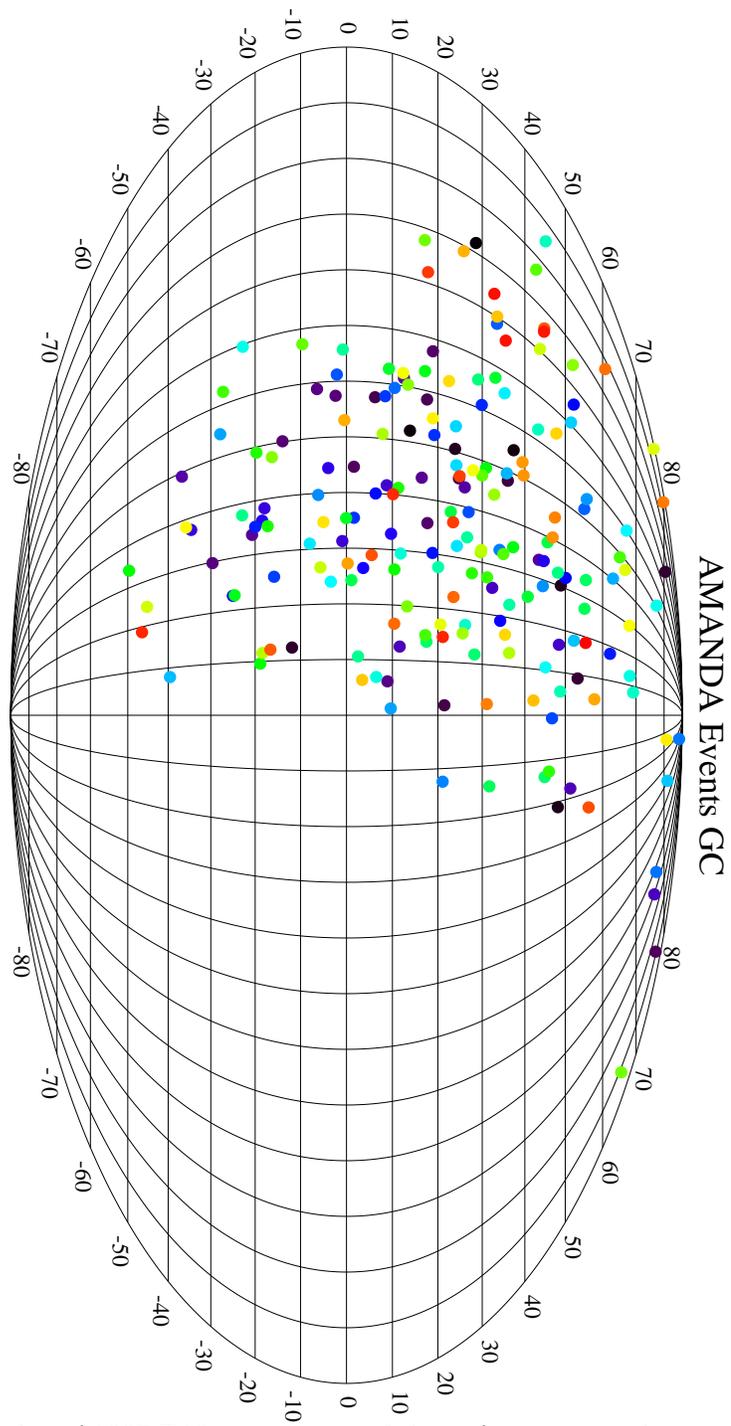, height=7.0in}
\caption{\label{skyplot} Skyplot of 1997 B10 neutrino candidates from one analysis in Galactic coordinates.  The detector has a significant vertical bias which is further exaggerated by quality criteria.}
\end{center}
\end{figure}

About $5\%$ of downgoing muons are mistaken for upgoing by the reconstruction, and quality criteria are necessary to identify these events automatically.  An effective cut parameter is $N_{\rm direct}$, the number of OM responses presumed to be minimally scattered photons from the $\check{\rm C}$erenkov cone of a given postulated track, to a specified degree of directness.  An event with many direct hits is one with a comparatively high content of uncorrupted information.  The number of direct hits is classified by $N_{\rm dirA}$, $N_{\rm dirB}$, and $N_{\rm dirC}$ to distinguish OM responses which occurred within $[-15,25]$, $[-15,50]$, and $[-15,75]$ nanoseconds of the expected time.  Combining these with various topological cuts improves up-down discrimination (Figure~\ref{zenith}) and angular resolution to acceptable levels.  Figure~\ref{skyplot} shows the angular distribution of the remaining neutrino candidates in the 1997 data passing a full set of quality cuts~\cite{ty}.

\chapter{Neutrinos from hell}
\label{flux}

\begin{quote}
\raggedleft {\em I am fire and air; my other elements I leave for the baser forms of life.} \\ William Shakespeare, {\em Anthony and Cleopatra}
\end{quote}

Gamma-ray bursts can be regarded as the relativistic analogue of supernova remnants (Table~\ref{snr_table}).  A highly efficient, relativistic outflow produces gamma radiation in internal shocks in the case of GRBs, whereas the non-relativistic outflow from supernovae produces optical emission in external shocks with notoriously low efficiency.
\begin{table}
\caption{\label{snr_table} GRBs are the relativistic analogue of supernova remnants.}
\begin{center}
\begin{scriptsizetabular}{|c c|} \hline\hline
{\bfseries SNR} & {\bfseries GRB}\\ \hline\hline
\multicolumn{2}{|c|}{Optical depth}\\
{\itshape high} & {\itshape low}\\ \hline
\multicolumn{2}{|c|}{Outflow velocity}\\
{\itshape non-relativistic} ($\Gamma<1.02$) & \mbox{\hspace{0.3cm}{\itshape relativistic} ($\Gamma>100$)\hspace{0.3cm}}\\ \hline
\multicolumn{2}{|c|}{Efficiency}\\
{\itshape low} & {\itshape high}\\ \hline
\multicolumn{2}{|c|}{Emission}\\
{\itshape optical} & {\itshape gamma rays}\\ \hline
\multicolumn{2}{|c|}{Shocks}\\
{\itshape external} & {\itshape internal}\\ \hline
\end{scriptsizetabular}
\end{center}
\end{table}
The analogy can be taken a step further:  just as supernovae are invoked to explain cosmic-ray energies up to $10^{15}$ eV and even beyond, GRBs represent a prime candidate location for cosmic-ray acceleration to energies beyond the ``ankle,'' the flattening in their spectrum near $10^{19}$ eV.  A compelling fact is that the inferred energy injection in gammas by GRBs, about $10^{44}$ erg Mpc$^{-3}$ yr$^{-1}$, is strikingly similar to that needed to produce the ultra-high energy cosmic rays (UHECRs)~\cite{wax2}.
\section{Hadronic GRBs}
The mass limit in Equation~\ref{mass_limit} is often referred to as the ``Baryon Contamination'' problem because it is difficult to develop a scenario in which only tiny quantities of material are entrained in the outflow of the fireball.  A small amount of baryons is actually useful in explaining how the fireball evolves from radiation-dominated to matter-dominated {\em before} becoming optically thin (Section~\ref{balls}).  

In portions of the flow where electrons are accelerated, protons could undergo shock acceleration as well, possibly to ultra-high energies.  Because the particles spend so little time on the shock front, it is difficult to efficiently accelerate protons on the highly relativistic external shocks.  Internal shocks, on the other hand, are expected to be only mildly relativistic, since Lorentz factors of different shells should differ by factors of order unity and not more than an order of magnitude.  If a sufficient fraction of the wind energy is carried by magnetic fields which can isotropize proton velocities on either side of the front, and if the proton acceleration time is shorter than both its synchrotron loss time and the wind expansion time, GRB fireballs could accelerate protons to $>10^{20}$ eV~\cite{wax1, vietri}.  There is now evidence for TeV~\cite{milagro} and sub-TeV~\cite{grand} gamma emission from GRBs which cannot be explained by a simple extrapolation of the BATSE energy spectra, but could be the result of $\pi_0$ production or synchrotron emission by energetic protons~\cite{grand}.
\section{The $\nu$ flux}
Protons accelerated in GRBs will photo-produce pions in interactions with fireball photons and the decay of these pions could generate a burst of high-energy neutrinos~\cite{bahc, jacz}:\\
\begin{tabular}{l}
\hspace*{8em}$p+\gamma\longrightarrow\Delta^+$\\
\hspace*{13.5em}$\searrow$\\
\hspace*{14.8em}$\pi^++n$\\
\hspace*{15.9em}$\searrow$\\
\hspace*{17.3em}$\mu^++\nu_\mu$\\
\hspace*{18.3em}$\searrow$\\
\hspace*{19.7em}$e^++\bar{\nu}_\mu+\nu_{\rm e}$\\
\end{tabular}
\newline
\newline The predicted energy distribution of particles accelerated on strong shocks is a power law with $N_{\rm p}(E_{\rm p})dE_{\rm p}\propto E_{\rm p}^{-(2+\epsilon)}dE_{\rm p}$.  These will interact with GRB photons to produce neutrinos which trace the broken power law of the seed photons:
\[N_\nu(E_\nu)dE_\nu\propto \frac{E_\nu^{-\beta}}{E_{\rm break}}dE_\nu\mbox{, with } 
\beta = \left\{ \begin{array}
{r@{\quad\mbox{for}\quad}l}
1 & E_\nu<E_{\rm break} \\ 2 & E_\nu>E_{\rm break}
\end{array} \right.\]
and
\begin{equation}
\label{break_eqn}
E_{\rm break}\approx7\times10^{14}\Gamma_{300}^2 \mbox{ eV,}
\end{equation}
where $\Gamma_{300}$ assumes a fireball Lorentz factor of 300.

Most calculations take advantage of the $\Delta$-$approximation$, which assumes the photo-production cross-section is dominated by the $\Delta$-resonance near threshold.  At high energies, non-resonant interactions of protons with the flat part of the photon spectrum could become significant~\cite{muck}.  Any neutrino emission is probably suppressed above about $10^{16}$ eV anyway, however, when the lifetimes of the intermediary muons and pions within the cascade exceed their synchrotron and adiabatic loss times.

The neutrino flux is determined by the factor $f_\pi\dot{\rho}$, where $\dot{\rho}$ is the injection rate of high-energy fireball protons and $f_\pi$ is the fraction of their energy given to pion production.  If GRBs are responsible for the UHECRs this fixes $\dot{\rho}\approx10^{44}$ erg Mpc$^{-3}$ yr$^{-1}$.  Taking $f_\pi=0.15$ the energy of GRB neutrinos can be normalized,
\[\mathcal{F}_\nu^{\rm GRB}=\frac{A}{E_{\rm break}}\int^{E_{\rm break}}_{E_{\rm min}}dE+A\int^{E_{\rm max}}_{E_{\rm break}}\frac{dE}{E}\mbox{,}\]
by determining the constant $A$, found by Halzen and Hooper~\cite{hoop} to be $1.2\times10^{-12}$ TeV cm$^{-2}$ s$^{-1}$ sr$^{-1}$.  Assuming 1000 bursts per year over $4\pi$ steridians, the fluence of neutrinos from a single GRB, in the compact form of Waxman and Bahcall, is then
\begin{equation}
\label{flux_per_burst}
E_{\nu}^2\frac{dN_\nu}{dE_\nu}=4.8\times10^{-7} {\rm min}\{1,E_\nu/E_{\rm break}\}\mbox{ TeV cm}^{-2}\mbox{.}
\end{equation}
This is consistent with the neutrino fluence estimated directly from the gamma rays~\cite{bahc2}:
\[E_{\nu}^2\frac{dN_\nu}{dE_\nu}=8.5\times10^{-7}(\frac{f_\pi}{f_e}) \mbox{ TeV cm}^{-2}\mbox{,}\]
where $f_e$ is the ratio of the energies of the electromagnetic and hadronic components of the fireball.

\begin{table}
\caption{\label{A_table} $A$ values for various models.  In the Halzen/Hooper and Waxman/Bahcall predictions burst parameters are fixed at typical values.  The model of Halzen, Alvarez and Hooper includes fluctuations in energy and distance.}
\begin{center}
\begin{scriptsizetabular}{|c||c|c|} \hline\hline
Model & $A$ & $E_{\rm break}$\\ \hline\hline
Halzen and Hooper & $1.2\times 10^{-12}$ & $700$ TeV\\
$\Gamma=300$ & &\\ \hline
Waxman and Bahcall & $4.5\times 10^{-12}$ & $500$ TeV\\
$\Gamma=300$ & &\\ \hline
Halzen, Alvarez and Hooper & $6.2\times 10^{-12}$ & $700$ TeV\\
$\Gamma=300$ & &\\ \hline
Halzen, Alvarez and Hooper & $9.1\times 10^{-11}$ & $77$ TeV\\
$\Gamma=100$ & &\\ \hline\hline
\end{scriptsizetabular}
\end{center}
\end{table}
Event rates in a neutrino detector can then be estimated by a convolution of the flux and detection probability of muon-flavored neutrinos, $P_{\nu\rightarrow\mu}$, determined by the interaction cross-section and muon range:
\[N=\int\phi_\nu(E)P_{\nu\rightarrow\mu}(E)dE\mbox{.}\]
This predicts that a detector with $1$ km$^2$ effective area should see on the order of $10$ events per year, which is discouragingly small.

It is heartening to note that this flux is the prediction for the {\em typical} GRB.  It may be far more likely to detect neutrino emission from a single burst with highly favorable properties than many standard-candle bursts.  An exceptionally close and powerful burst could of course provide better-than-average observational possibilities.  Fireball parameters are not well constrained and the photon target density and hence the value of $f_\pi$ and the break energy all depend sensitively on the Lorentz factor $\Gamma$ of the outflow.  Burst-to-burst fluctuations could increase the neutrino intensity from a particular burst, perhaps dramatically higher than that in Equation~\ref{flux_per_burst}~\cite{hoop, hah}.  Some effective $A$ values, which take into account fluctuations in energy and distance, are shown in Table~\ref{A_table}.  Integrating the flux of GRB neutrinos from $E_{\rm min}$, the $1.232$ GeV threshold of the $\Delta$-resonance,  the fluence is then proportional to
\[E_{\nu}^2\frac{dN_\nu}{dE_\nu}\propto\frac{A}{E_{\rm break}}(\ln{\frac{E_{\rm break}}{E_{\rm min}}}+1)\mbox{.}\]
More detailed observation will constrain the degree to which parameters of the GRB fireball fluctuate and how those fluctuations could be manifest in any neutrino emission.  
\subsection{Scientific goals}
A measurement of the GRB neutrino flux can test the hypothesis of GRBs as the origin of the highest energy cosmic rays.  Like SN1987A, GRBs could permit even stronger tests of the neutrino limiting speed and the weak equivalence principle at levels of $10^{-16}$ and $10^{-6}$, respectively, due to their short duration.  Furthermore, since no $\nu_\tau$ are expected to be produced in the source, an appearance of $\nu_\tau$ would be clear evidence of neutrino oscillation at a value of ${\Delta}m^2$ as low as $10^{-16}$ eV$^2$.
\subsection{Monte Carlo simulation}
\label{shadow}
\begin{figure}[h]
\begin{center}
\epsfig{file=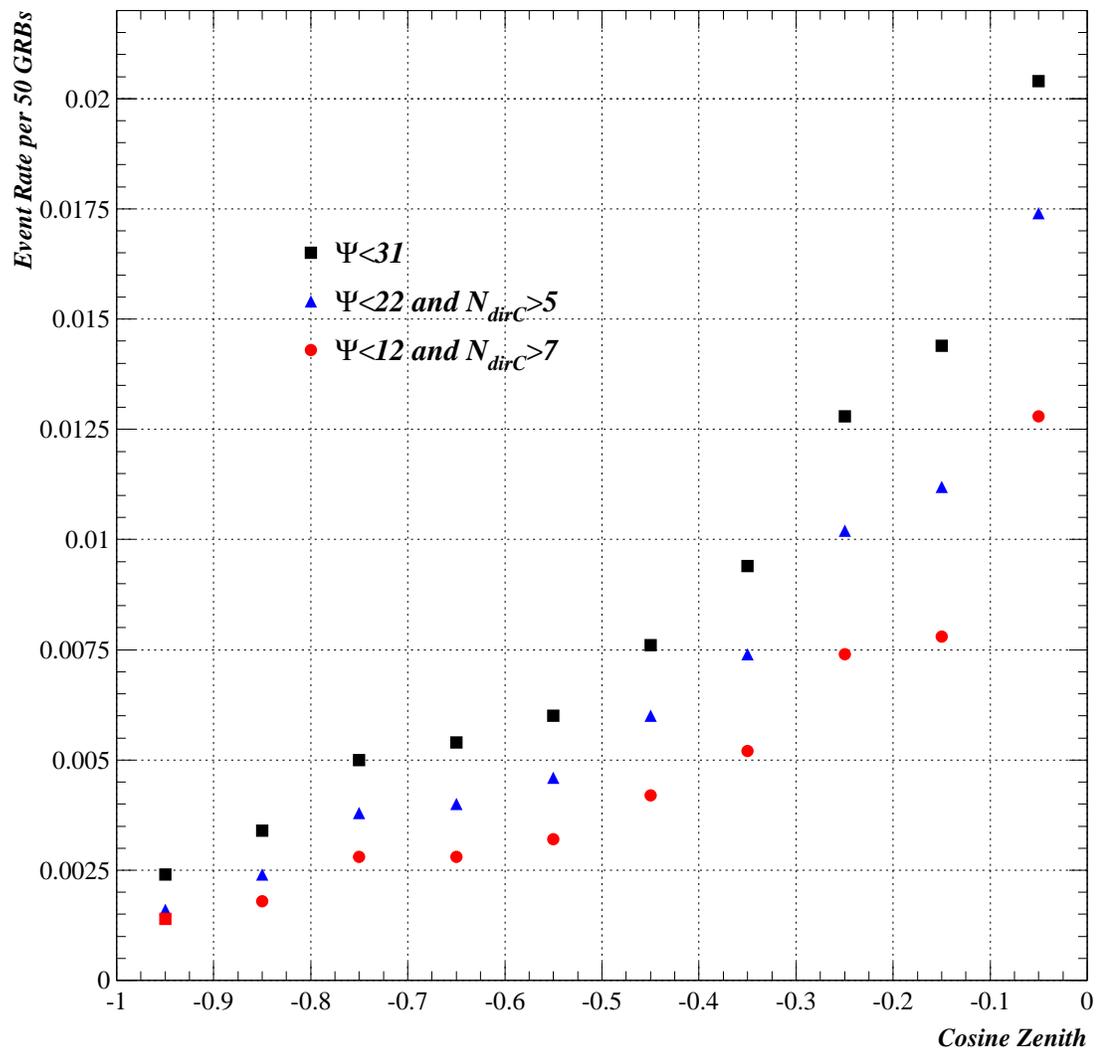, height=5.5in}
\caption{\label{shadow_fig} Monte Carlo calculation gives expected event rates from 500 GRBs spread over 10 zenith bins.  The dependence is dominated by neutrino absorption in the Earth.}
\end{center}
\end{figure}

The stiff GRB neutrino flux should suffer significant Earth shadowing~\cite{gandhi}.  Monte Carlo simulation~\cite{hill} gives the expected event rate at three cut levels (Section~\ref{optimum}) in AMANDA-B10 versus zenith angle for 500 Halzen and Hooper GRBs distributed over $2\pi$ steridians in Figure~\ref{shadow_fig}.  The sensitivity of the array drops by nearly an order of magnitude near nadir, where effective area and angular resolution are otherwise optimum.

\chapter{Data processing}

\section{Data collection}
\subsection{GRB data}
\setcounter{footnote}{1}
Localized burst data are available from three sources, and I compiled triggers in the following priority:\\  
\hspace*{2cm}i) BeppoSAX triggers,\\
\hspace*{2cm}ii) bursts detected by three or more Inter-Planetary Network (IPN) satellites, and\\
\hspace*{2cm}iii) BATSE triggers.\\
The BeppoSAX Wide Field Camera covers about $5\%$ of the sky and together with the Narrow Field Instrument provides arcminute locations to a few bursts per year.  Bursts detected by satellites in the IPN must be observed by at least three widely separated spacecraft for their position to be well localized.  While there are many spacecraft near Earth, in interplanetary space there were only two in 1997, Ulysses and the Near Earth Asteroid Rendezvous (NEAR).  BATSE remains the primary source for GRB triggers as of this writing\footnote{BATSE and the other instruments on-board the Compton Gamma-Ray Observatory were turned off at UT midnight on my 28$^{th}$ birthday, May 26, 2000, due to fears of loss of spacecraft gyro-control.  Compton-GRO was removed from orbit and burned up in the Earth's atmosphere on June 4, 2000.}.  BATSE triggers are based mostly on counts in the $50$ to $300$ keV range and it is very possible that a large population of hard GRBs escapes detection.  Searches of BATSE data have produced many hundreds of non-triggered burst candidate events~\cite{boris}, most localized to no better than $10^\circ$.  Some events have been confirmed by the Ulysses experiment, while some others are probably spurious~\cite{kevin}.

BATSE consists of eight modules of NaI(Tl) scintillation detectors, one module on each corner of the Compton-GRO satellite.  BATSE can estimate source direction by the relative intensities in the eight modules and the BATSE team reports the statistical error radius of a $68\%$ confidence ellipse.  For a weak burst, the positional error can be as large as $20^\circ$, but for strong bursts it is less than one degree.  There is in addition a systematic error of about $1.6^\circ$($73\%$ probability)~\cite{batse}.  The BATSE team uses $T_{90}$($T_{50}$), the time required to accumulate $5\%(25\%)$ to $95\%(75\%)$ of the total counts, to characterize burst duration.  Those bursts which suffer data gaps during the event (about $15\%$ of the triggers in 1997) are not assigned $T_{90}$($T_{50}$) times, although some visual estimates of duration are available in the BATSE Comments Table~\cite{batse}.

\subsection{AMANDA data}
I extracted all two-hour intervals of AMANDA-B10 muon data bracketing 1997 GRBs from the collaboration repository after the Level 1 filter.  These events have been calibrated and reconstructed with both a simple line fit and the full reconstruction, while ignoring malfunctioning optical modules and isolated hits presumed to be noise.  The line fit result is required to have $\theta>50^\circ$ for an event to be part of the $22\%$ passing this stage of the filter~\cite{jodi}.

\begin{figure}[p]
\begin{center}
\epsfig{file=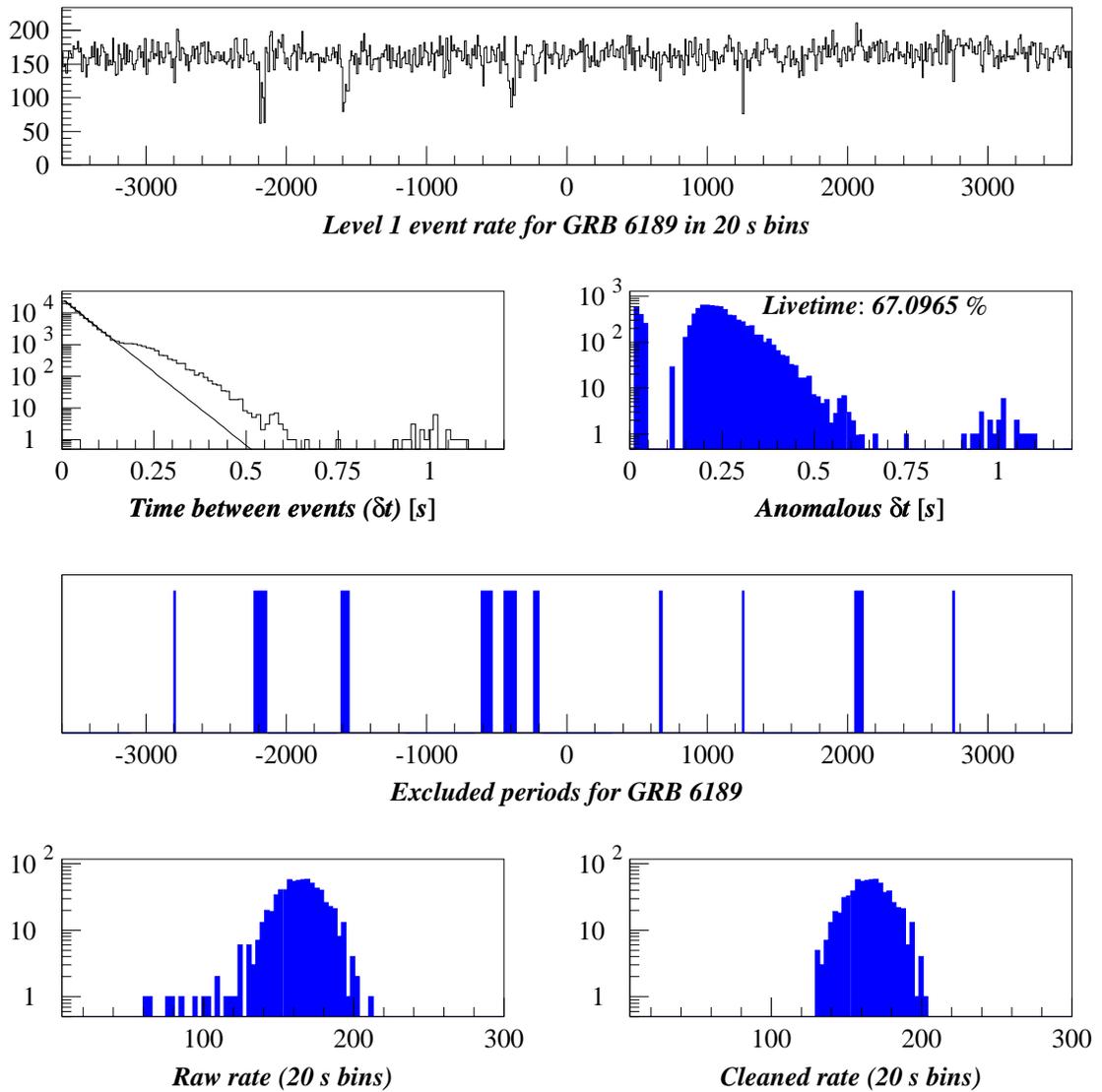, height=5.8in}
\caption{\label{exclude}The cleaning algorithm results during a somewhat erratic period of AMANDA behavior.  Event gaps introduced by the data acquisition system which exceed a fit to the natural B10 trigger exponential were integrated to determine live time, which averaged about 68\% in 1997.  Time periods with highly improbable total array rates or excessive gaps were excluded from the analysis.}
\end{center}
\end{figure}

I first analyzed these data to remove periods of uncharacteristic detector performance from the search, isolating hardware and software glitches by consulting previous data diagnostics~\cite{ped} and using a specifically designed algorithm.  Each two-hour measurement was divided into ten-second intervals and evaluated on the basis of total array rate and occurrence of long gaps between events.  Time bins in which the total array rate deviated by more than $3.5$ to $4$ standard deviations from a running average or which contained more than two gaps longer than $0.5$ seconds were removed.  Bins immediately next to a failing bin were required to pass even stronger criteria.  The cuts were tuned by visual inspection to minimize biasing.  Figure~\ref{exclude} shows an example application of the algorithm and a calculation of live time.  During a few bursts the muon data were completely unusable.

\begin{figure}[h]
\begin{center}
\epsfig{file=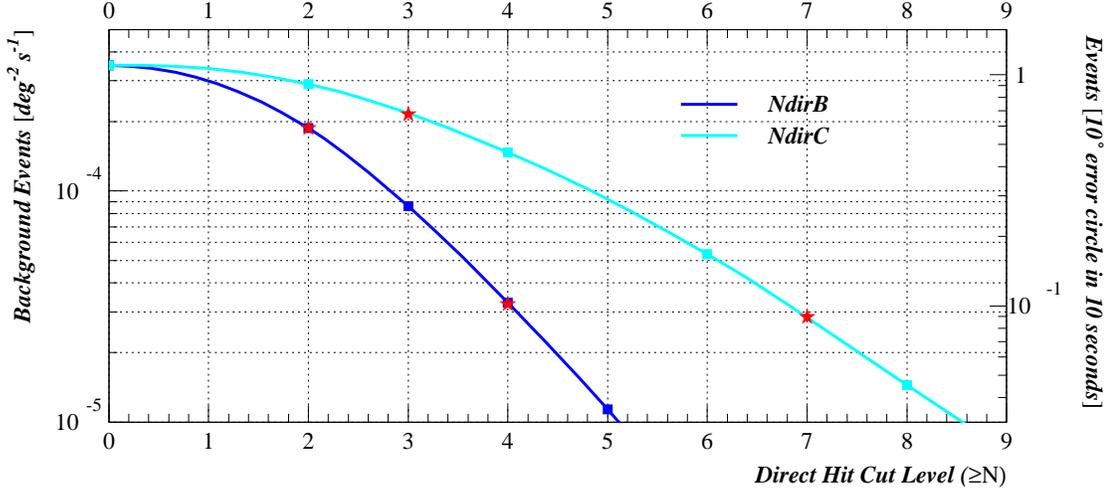, height=2.6in}
\caption{\label{bkg_removal} Misreconstructed upgoing background at Level 1, both per degree$^2$ per second and in a $10$ degree error circle in $10$ seconds.}
\end{center}
\end{figure}

Figure~\ref{bkg_removal} shows how the background of noise and misreconstructed downgoing events is removed as more direct hits are required in the events.  The cut $N_{\rm dirB}\ge2$ results in roughly the same background suppression as $N_{\rm dirC}\ge3$, and $N_{\rm dirB}\ge4$ is comparable to $N_{\rm dirC}\ge7$, yet the $N_{\rm dirC}$ cut preserves more signal than the $N_{\rm dirB}$ cuts in each case (Figure~\ref{point}).

\section{Optimization}
\label{optimum}
\begin{figure}[h]
\begin{center}
\epsfig{file=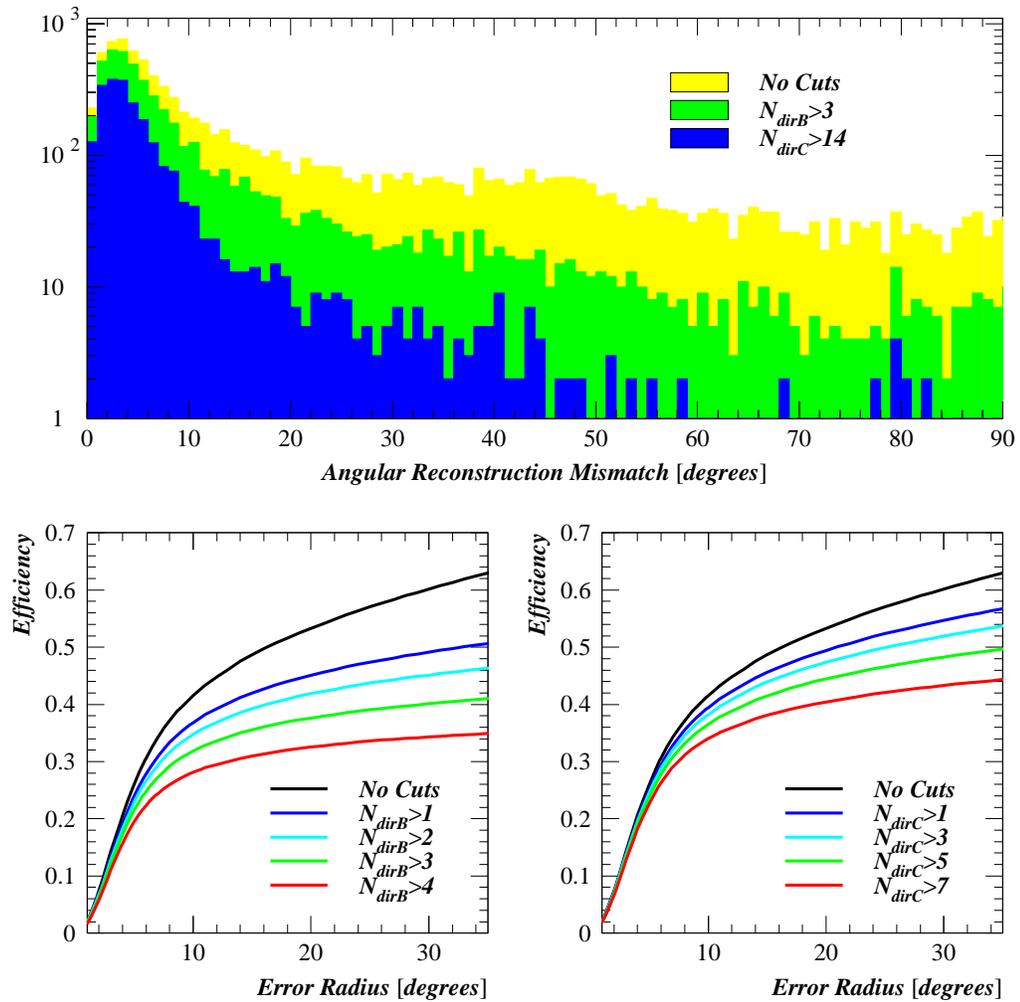, height=5.6in}
\caption{\label{point} {\em Top}:  Monte Carlo simulation of the point spread function of the GRB neutrino spectrum in AMANDA-B10.  {\em Bottom}:  The efficiency as a function of error circle radius at various cut levels.}
\end{center}
\end{figure}

\begin{figure}[p]
\begin{center}
\epsfig{file=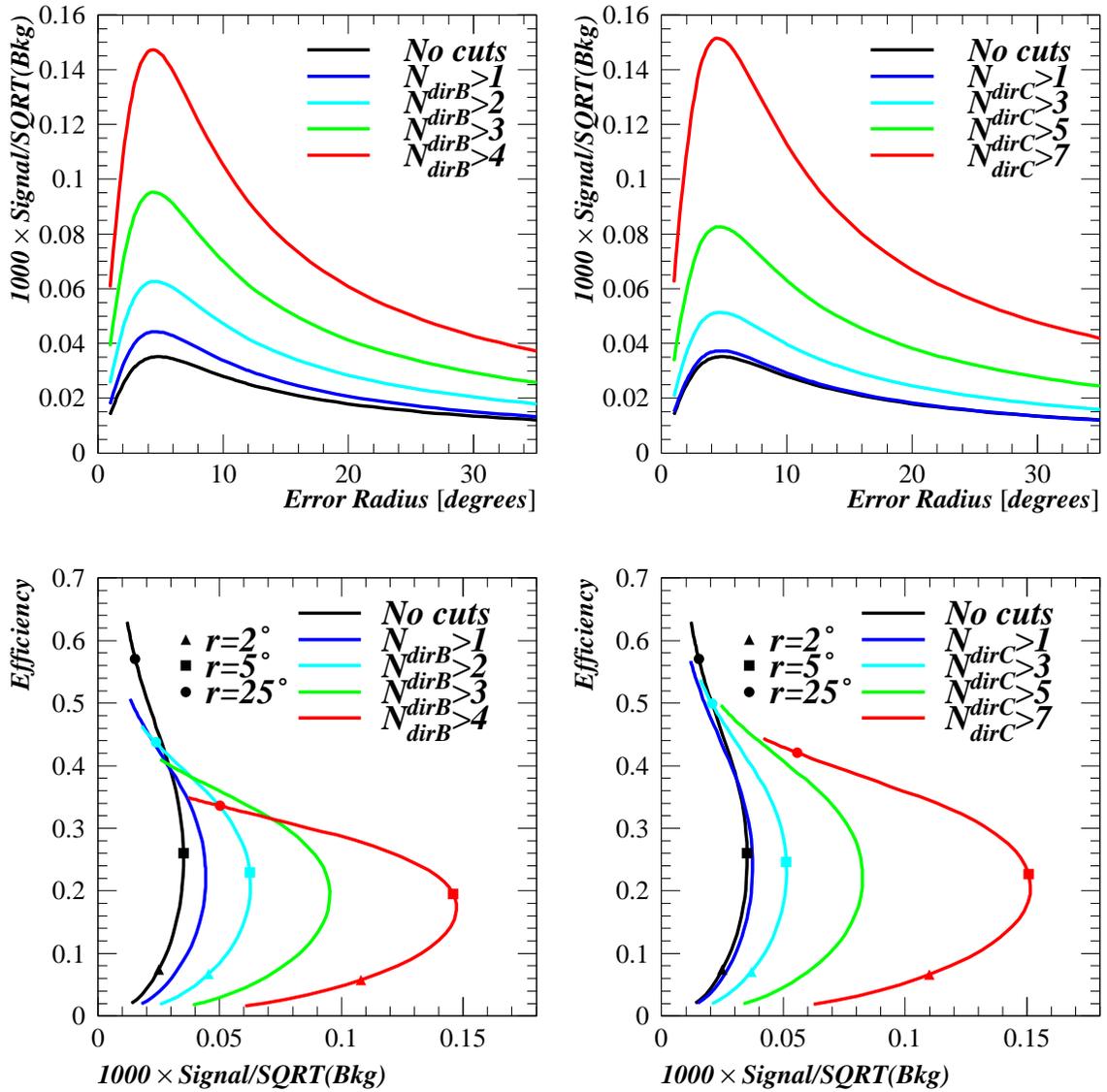, height=5.9in}
\caption{\label{opt} {\em Top}:  Q-factor ($signal/\sqrt{background}$) versus search bin size.  {\em Bottom}:  Efficiency versus Q-factor.}
\end{center}
\end{figure}

Neutrino candidate searches in the current AMANDA typically employ a battery of event quality criteria which suppresses signal by $99\%$~\cite{karle}.  The transient and localized nature of GRBs allows quality criteria to be relaxed and the sensitivity greatly increased.  I optimized cuts using a Monte Carlo simulation~\cite{hill} of the response of AMANDA-B10 to the canonical GRB neutrino spectrum, and AMANDA-B10 data for background.  Muon propagation was simulated using MUDEDX~\cite{sieg} software and the detector response using AMASIM~\cite{hundert}.  The point spread function of reconstructed GRB events is shown in the top part of Figure~\ref{point}.  The bottom figure shows how the efficiency increases as the error circle is enlarged, with gains in signal diminishing after events accumulate quickly in the first several degrees.  Figure~\ref{opt} shows how the Q-factor ($signal/\sqrt{background}$) and efficiency vary together with search bin size.  The events were generated from a constant direction near $\theta=45^\circ$, $\phi=135^\circ$.  At high energies, angular dependence of detector resolution is not expected to be dramatic.
\vspace*{1.0cm}  

\newpage
\setcounter{footnote}{1}
Optimum cuts are those that perform in the uppermost and rightmost parts of the graph of efficiency versus Q-factor on the bottom of the Figure~\ref{opt}.  I chose the three cocktails\footnote{In the first analysis of the data~\cite{ryan}, I chose $N_{\rm dirB}>4$, $\Psi<10^\circ$, which is slightly inferior.} in Table~\ref{cut_table}.  The tightest cut provides the best discrimination while retaining a statistically meaningful number of events and yields the greatest gain when combining measurements to improve an upper limit due to the low background, with $L\propto\frac{1}{N_{\rm meas}}$ as opposed to $L\propto\frac{1}{\sqrt{N_{\rm meas}}}$ with high background.  The Q-factor could be further optimized with more sophisticated cuts on Monte Carlo simulations of a better-constrained GRB neutrino spectrum, and by making independent zenith and azimuth detector corrections.
\begin{table}
\caption{\label{cut_table}Selected cuts.}
\begin{center}
\begin{scriptsizetabular}{|c|c||c|c|c|} \hline\hline
N$_{\rm direct}$ cut & Bin radius($\Psi$) & Background (10 s) & Efficiency & Relative Q ($S/\sqrt{Bkg}$)\\ \hline\hline
None & $31^\circ$ & $\sim10$ & 0.61 & 1.0\\ \hline
$N_{\rm dirC}>5$ & $22^\circ$ & $\sim1$ & 0.45 & 2.7\\ \hline
$N_{\rm dirC}>7$ & $12^\circ$ & $\sim0.1$ & 0.36 & 7.5\\ \hline\hline
\end{scriptsizetabular}
\end{center}
\end{table}

\section{On-Off source event counting}
I took the satellite resolution for a given burst to be the reported error radius divided by $1.58$ after Alexandreas et al.~\cite{alex}, but did not employ this optimized scaling based on expected numbers of events to then enlarge the already substantial search bin radii equal to the quadrature sum of the angular cuts listed in Table~\ref{cut_table} and the satellite resolution:
\[\Psi_{search}=\sqrt{\Psi_{cut}^2+(\frac{\mbox{satellite position error}}{1.58})^2}\mbox{.}\]
Only GRBs within the Northern Hemisphere ($\theta>90^\circ$) were included in the search.  Hard cuts were made at the error circle and at $\theta=80^\circ$ and all events falling within an error bin were weighted equally.  

Since particle acceleration is expected to persist only during the internal shocks associated with the gamma-ray flash, I confined the search to roughly the duration of each GRB.  I tallied the number of events, $n_{0}$, falling in the angular search bin beginning $100$ milliseconds before and ending $100$ milliseconds after the $T_{90}$ time.  For those bursts for which duration information was not available, I searched over the interval [-2,8] seconds around the trigger time.  

I estimated background by averaging events in the same detector angular bin (fixed $\theta$ and $\phi$) during all other intervals which passed the quality analysis.  For most bursts the off-source observation was many hundreds or thousands of times the $T90$ interval and the error in the number of expected events, $\mu_{\rm B}$, is negligible.  

Results and analysis appear in the next chapter.

\chapter{Results and conclusions}
\label{conc}

\begin{quote}
\raggedleft {\em If the glove don't fit, you must acquit!} \\ Johnnie Cochran
\end{quote}

\section{Search results}
\begin{figure}[h]
\begin{center}
\epsfig{file=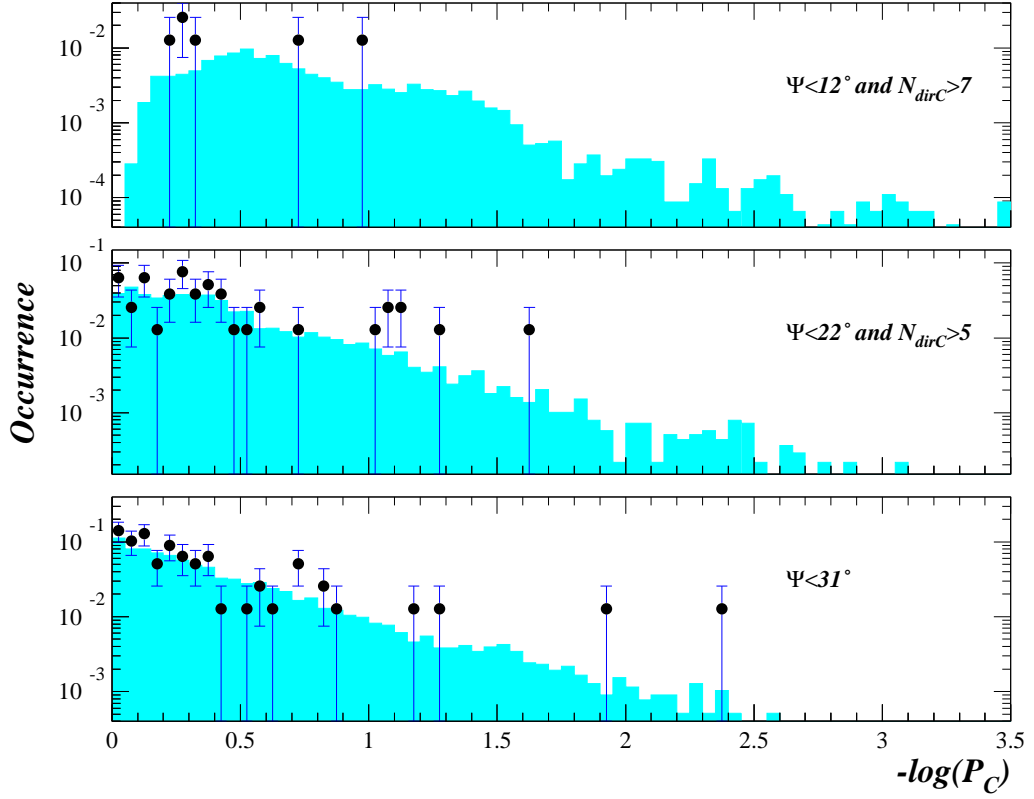, height=4.2in}
\caption{\label{chance}Chance probabilities, $P_{\rm C}$, of on-source searches are consistent at all cut levels with that expected with no signal (solid histogram).}
\end{center}
\end{figure}

Figure~\ref{chance} shows the results of the search for neutrino emission during the gamma-ray flash of 78 BATSE GRBs at the three cut levels.  I compare the chance probability of occurrence to that expected with no signal, determined by a Monte Carlo simulation of several thousand searches in random angular bins.  Results are consistent with no signal.  

\begin{figure}[p]
\begin{center}
\epsfig{file=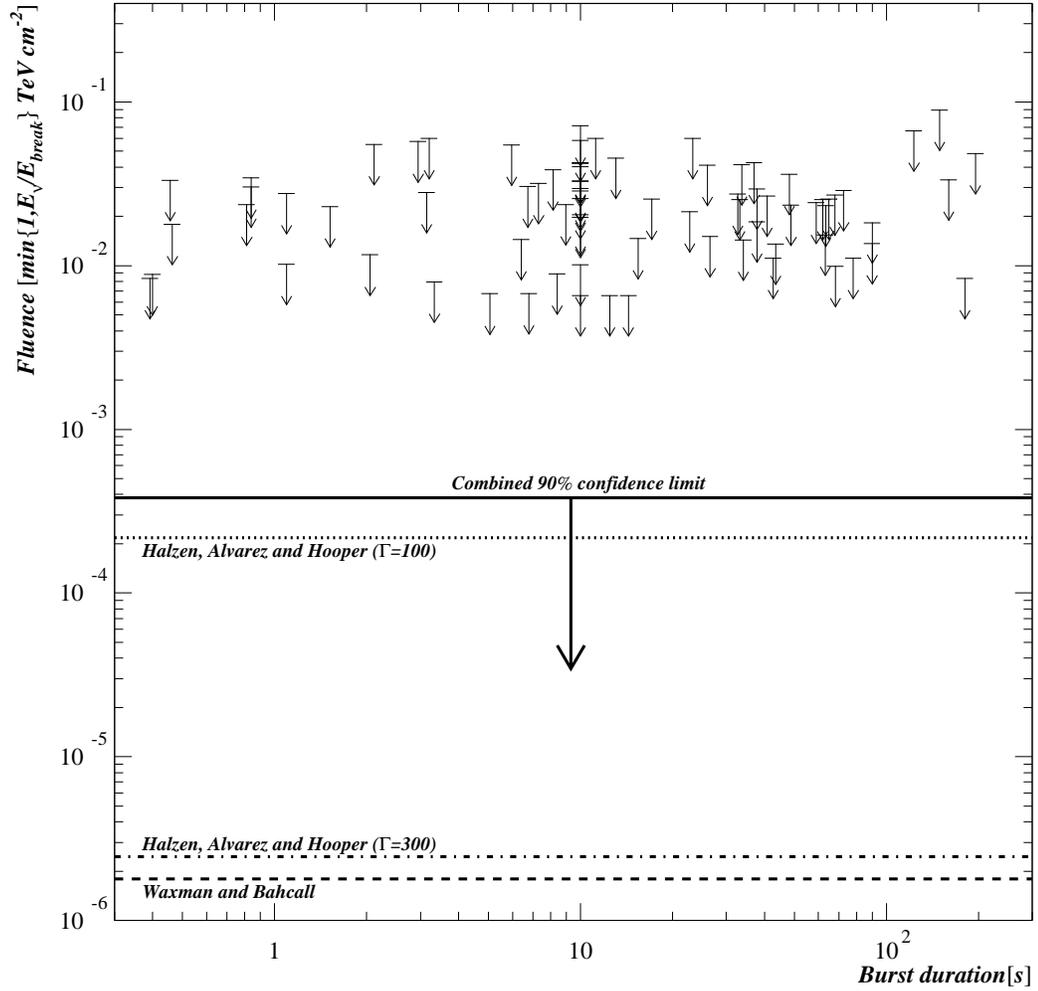, height=5.2in}
\caption{\label{limit7}Limits on neutrino fluence during the gamma-ray flash of 78 GRBs in 1997 versus burst duration at the tightest cut level.  The solid line is the combined fluence limit, while dashed lines are model predictions.}
\end{center}
\end{figure}

\begin{figure}[p]
\begin{center}
\epsfig{file=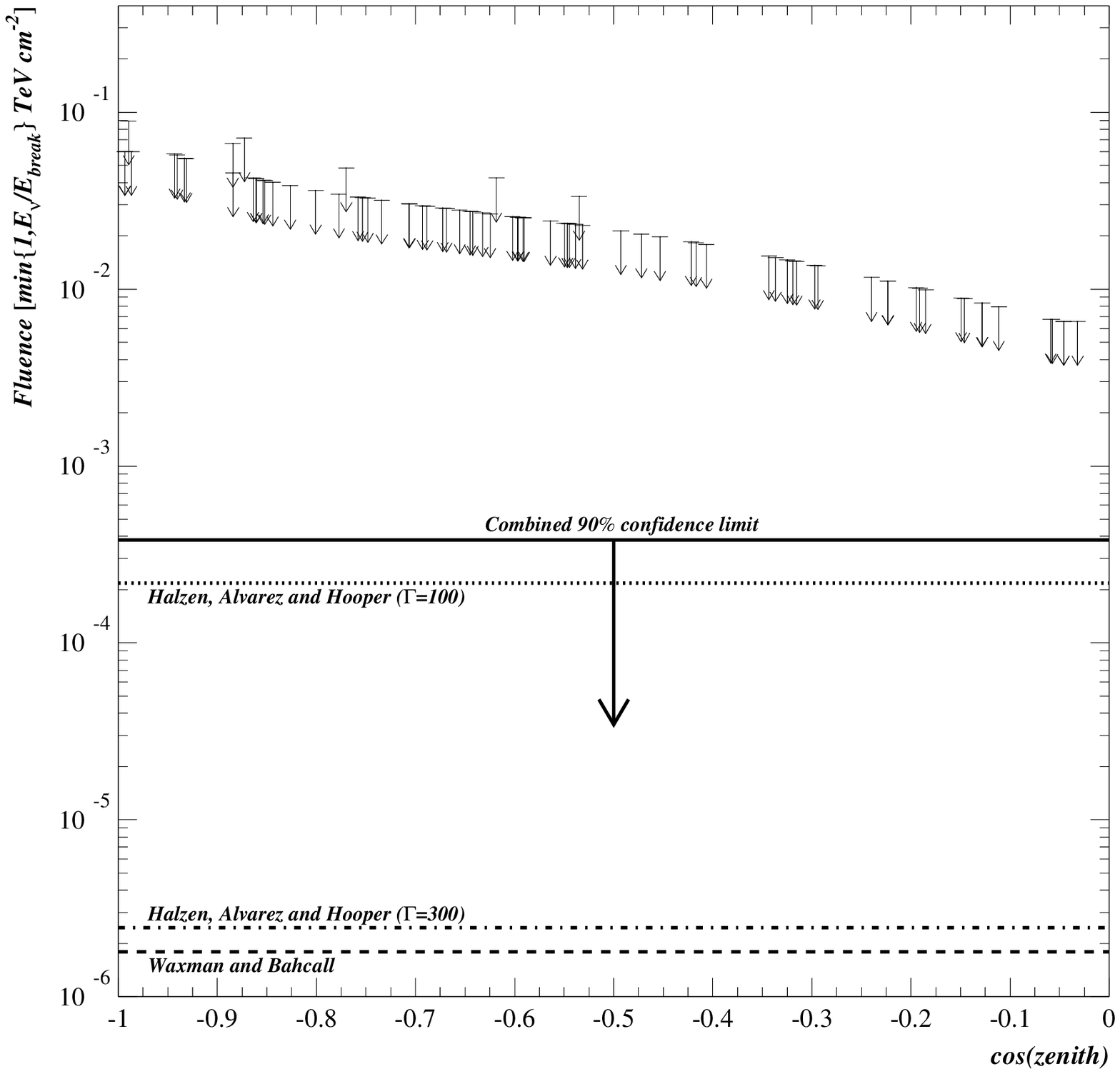, height=5.2in}
\caption{\label{zen7}Limits on neutrino fluence during the gamma-ray flash of 78 GRBs in 1997 versus $\cos(\theta)$ at the tightest cut level.  The solid line is the combined fluence limit, while dashed lines are model predictions.}
\end{center}
\end{figure}

\begin{figure}[p]
\begin{center}
\epsfig{file=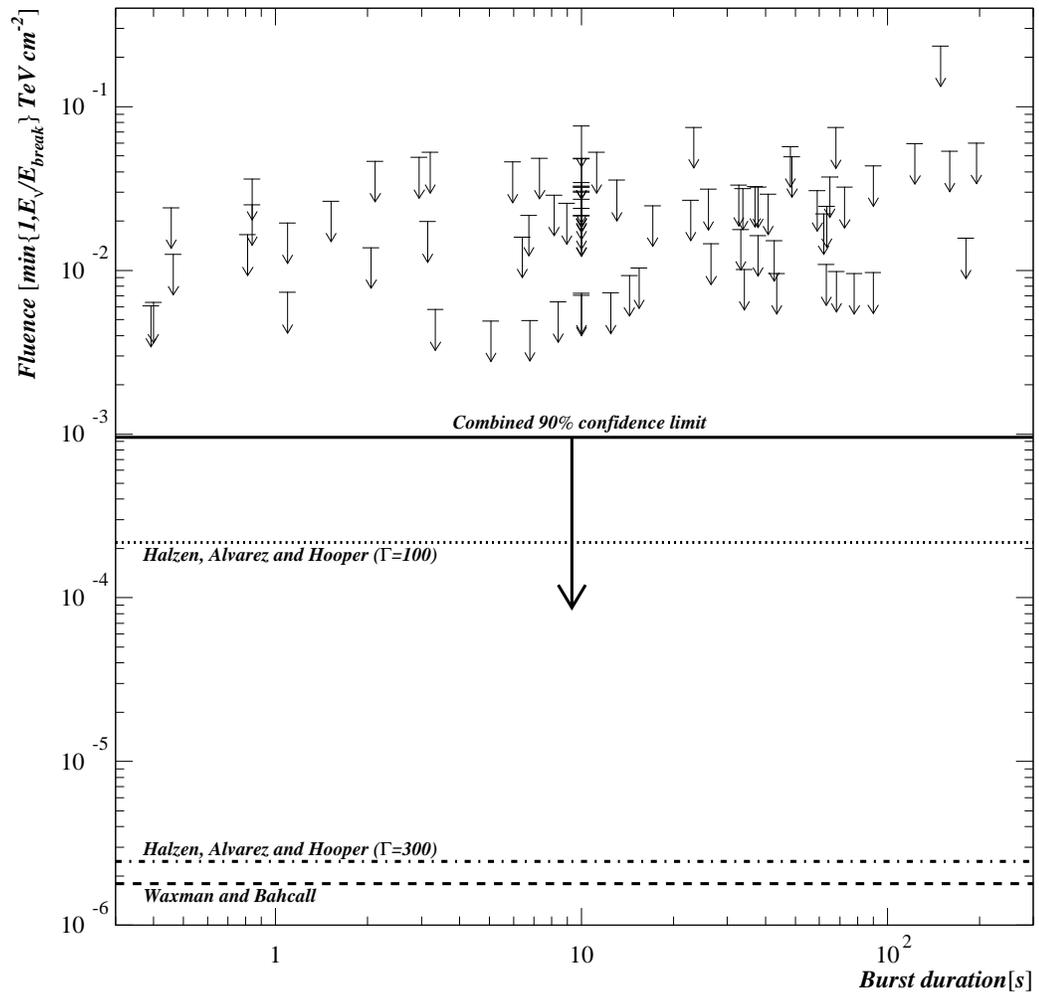, height=5.2in}
\caption{\label{limit5}Fluence limit versus duration at medium cut level.}
\end{center}
\end{figure}

\begin{figure}[p]
\begin{center}
\epsfig{file=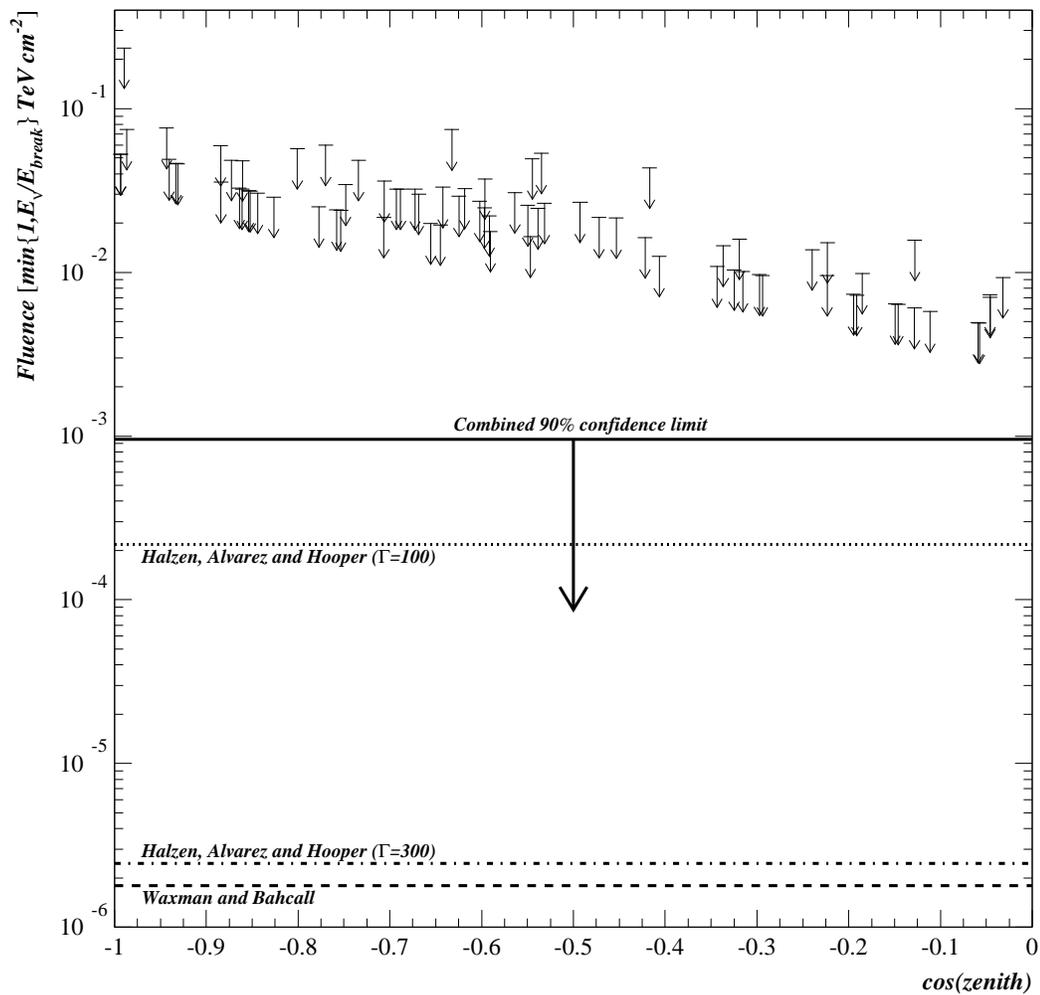, height=5.2in}
\caption{\label{zen5}Fluence limit versus $\cos(\theta)$ at medium cut level.}
\end{center}
\end{figure}

\begin{figure}[p]
\begin{center}
\epsfig{file=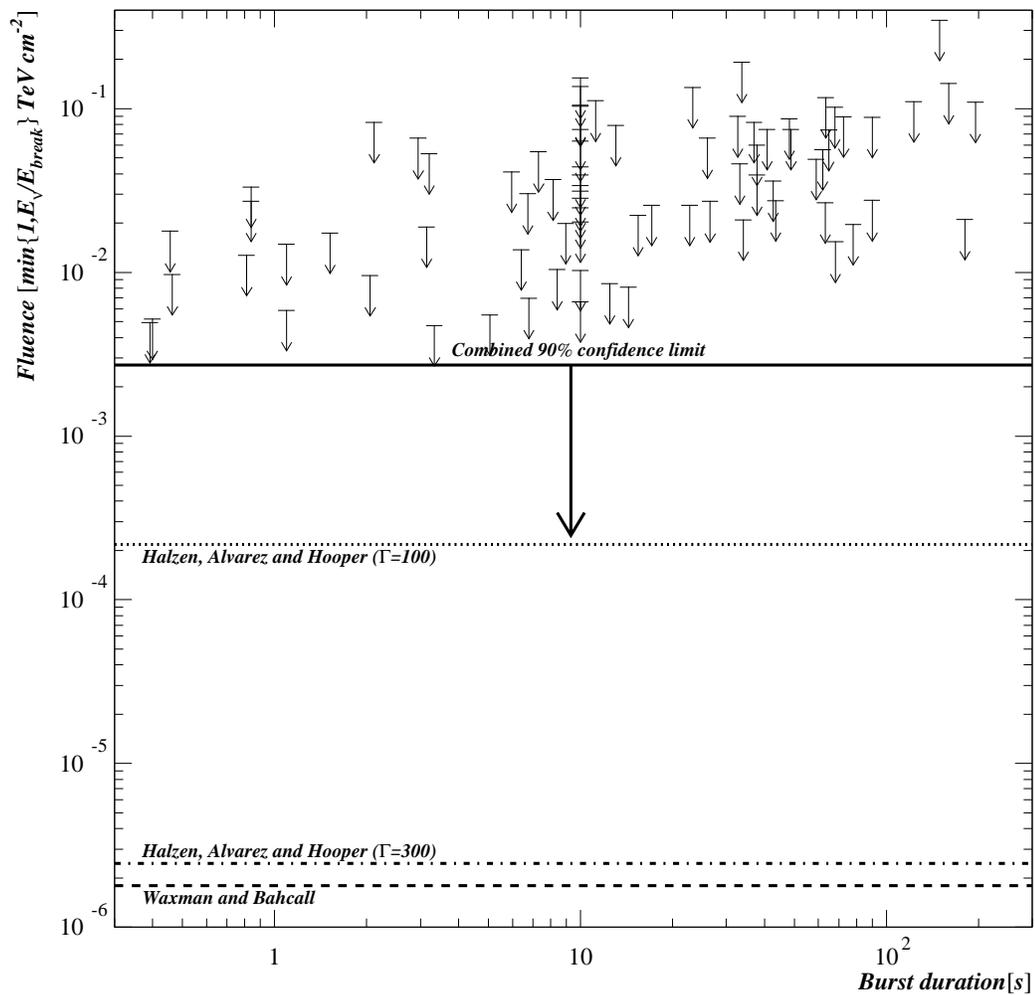, height=5.2in}
\caption{\label{limit0}Fluence limit versus duration at the loosest cut level.}
\end{center}
\end{figure}

\begin{figure}[p]
\begin{center}
\epsfig{file=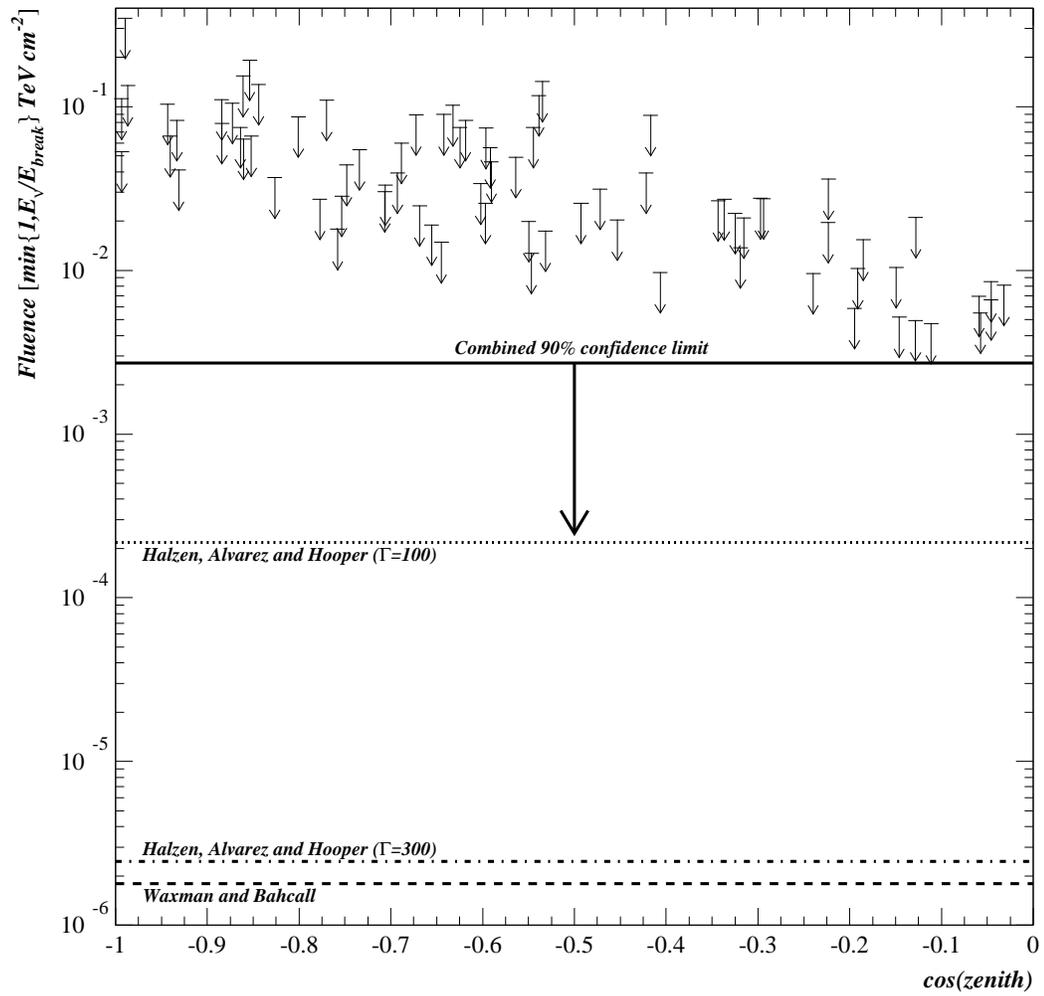, height=5.2in}
\caption{\label{zen0} Fluence limit versus $\cos(\theta)$ at the loosest cut level.}
\end{center}
\end{figure}

Employing the prescription in Montanet et al.~\cite{mont} (page 1281) for a Poisson process with background, I place a limit on signal events for each burst and for the combined sample.  The calculation determines the number of signal events, $N$, such that any random repetition of the experiment with the same expected background, $\mu_{\rm B}$, would observe more than the observed number of events $n_{0}$ and would have $n_{\rm B}\le n_{0}$, all with probability $90\%$:
\begin{equation}
\frac{e^{-(\mu_B+N)}\sum\limits_{n=0}^{n_0}\frac{(\mu_B+N)^n}{n!}}{e^{-\mu_B}\sum\limits_{n=0}^{n_0}\frac{\mu_B^n}{n!}}=0.9\mbox{.}
\end{equation}
%\begin{table}
%\caption{\label{events} On source, off source and signal event limits at the three cut levels.}
%\begin{center}
%\begin{scriptsizetabular}{|c||c|c|c|c|} \hline\hline
%Cut & $n_0$ & $\mu_B$ & $N$ & Halzen and Hooper fluence\\ \hline\hline
%$\Psi<12^\circ$,$N_{\rm dirC}\ge7$ & $6$ & $12.7$ & $3.7$ & $7.9\times10^2$\\ \hline
%$\Psi<22^\circ$,$N_{\rm dirC}\ge5$ & $121$ & $136.2$ & $12.9$ & $2.0\times10^3$\\ \hline
%$\Psi<31^\circ$ & $1676$ & $1722$ & $46.7$ & $5.7\times10^3$\\ \hline\hline
%\end{scriptsizetabular}
%\end{center}
%\end{table}
Limits are naturally strongest for short bursts near the horizon at the tightest cut level where the Q-factor is relatively high (Figures~\ref{limit7} through~\ref{zen0}).  Taking the response in AMANDA-B10 given by Monte Carlo simulation (Section~\ref{shadow}) and the measured live time of $68\%$, I convert the event limits into a neutrino fluence.  With the hard cut I find a combined $90\%$ confidence limit
\begin{equation}
\label{result}
E_{\nu}^2\frac{dN_\nu}{dE_\nu}<3.8\times10^{-4} {\rm min}\{1,E_\nu/E_{\rm break}\}\mbox{ TeV cm}^{-2}\mbox{,}
\end{equation}
per average burst during the gamma-ray flash.  This limit is orders of magnitude more stringent than the limits of $\sim10^{-8}$ and $\sim10^{-9}$ upward-going muons cm$^{-2}$ per burst set by the Irvine-Michigan-Brookhaven (IMB)~\cite{imb} and the Monopole Astrophysics and Cosmic Ray Observatory (MACRO)~\cite{macro} neutrino detectors.

\section{Outlook}
The fluence limit in Equation~\ref{result} is comparable to the prediction for $\Gamma=100$, and careful modeling of the fireball should permit even the current AMANDA to place a useful lower bound on the value of $\Gamma$.  Predictions of the GRB neutrino flux will improve as more is learned about conditions in GRB fireballs with the launch of a new generation of sophisticated satellites.  Future GRB missions or further analysis of old data can be expected to identify more GRBs, in particular the harder variety accompanied by the sub-TeV to TeV emission suggestive of particle acceleration.

At the preferred cut level used in this search, limits improve roughly as the number of bursts in the sample and the addition of the 1998 and 1999 data sets will further constrain neutrino emission by a factor of $3$ to $4$.  Enlarging the data set also improves the odds of observing the peculiar burst with favorable characteristics.  The likelihood of GRB neutrino detection is certain to increase as AMANDA and detectors like it continue to expand.

\bibliographystyle{prsty}
\bibliography{myuc}

%\appendix
%\chapter{Some Ancillary Stuff}

%Ancillary material should be put in appendices, which appear after the 
%bibliography. 

\end{document}